%Paper: hep-th/9411099
%From: a.tseytlin@ic.ac.uk
%Date: Mon, 14 Nov 94 20:54:47 gmt
%Date (revised): Wed, 14 Dec 94 21:46:05 gmt
%Date (revised): Tue, 11 Apr 95 10:19:48 bst
%Date (revised): Tue, 11 Apr 95 19:18:56 bst

 \input harvmac

\def \o{\omega}
\def\const {{\rm const}}
\def \s {\sigma}
\def\t {\tau}
\def \bt { \bar \tau}
\def \p {\phi}
\def \ha {\half}
\def \ov {\over}
\def \fourth  {{1\ov 4} }
\def \four {\textstyle {1\ov 4}}
\def \a {\alpha}
\def \lr { \lref}
\def\ep{\epsilon}

\def\vp {\varphi}
\def \bd {\bar \del}\def \B {\bar B}
\def \r {\rho}
\def\const {{\rm const}}\def\bd {\bar \del} \def\m{\mu}\def\n {\nu}\def\l
{\lambda}\def\hG {{\hat \Gamma}}\def \G {\Gamma}

\def \f {  f }
\def \E {{ E}}
\def \pp {{ p}}
\def\Tau{{\cal T}}
\def\g {\gamma}

\def \pp {{ p}}

 \def \sm {$\s$-model\ }
\def   \td {\tilde }
\def \k {\kappa}
\def \lr { \lref}

\gdef \jnl#1, #2, #3, 1#4#5#6{ { #1~}{ #2} (1#4#5#6) #3}

\lr \burg{C.P. Burgess, \np B294 (1987) 427; V.V. Nesterenko, \ijmp A4 (1989)
2627.}
\lr \tsenul {A.A. Tseytlin, \np B390 (1993) 153. }
\lr \susskind { L. Susskind, ``Some speculations about black hole entropy in
string theory", RU-93-44 (1993), hep-th/9309135. }

\lr\hrt {G.T. Horowitz and A.A. Tseytlin. \pr D50 (1994) 5204. }

\lr \gtwo {E. Del Giudice, P. Di Vecchia and S. Fubini, Ann. Phys. 70
(1972) 378; K. A. Friedman and C. Rosenzweig, Nuovo Cimento 10A (1972) 53;
S. Matsuda and T. Saido, Phys. Lett. B43 (1973) 123; M. Ademollo {\it et al},
Nuovo Cimento A21 (1974) 77;
S. Ferrara, M. Porrati and V.L. Teledgi, Phys. Rev. D46
(1992) 3529.}

\lr \plane { D. Amati and C. Klim\v c\'\i k,
\jnl \pl, B219, 443, 1989;
 G. Horowitz and A. Steif,  \jnl \prl, 64, 260, 1990; \jnl \pr,
D42, 1950, 1990;
 G. Horowitz, in: {\it
 Strings '90}, eds. R Arnowitt et. al.
 (World Scientific, Singapore, 1991);
 H. de Vega and N. S\' anchez, \pr D45 (1992) 2783; Class. Quant. Grav. 10
(1993) 2007.}

\lr \gibma { G.W.  Gibbons and  K. Maeda, \np B298 (1988) 741;
G.W.  Gibbons, in: {\it Fields and Geometry}, Proceedings of the 22-nd Karpacz
Winter School of Theoretical Physics, ed. A. Jadczyk (World Scientific,
Singapore,  1986).}

\lref \tsnul { A. Tseytlin, \jnl \np, B390, 153, 1993.}

\lr\duval  { C. Duval, Z. Horvath and P.A. Horvathy, \jnl \pl,  B313, 10,
1993.}

\lr\gro{D.J.  Gross, J.A. Harvey,  E. Martinec and R. Rohm, \np B256 (1985)
253; \np B267 (1985) 75.}
\lr \duff {Duff et al Nepomechie }
\lr \incer { E.J. Ferrer, E.S. Fradkin and V.de la Incera, \pl B248 (1990)
281.}
\lr \green {M.B. Green, J.H.  Schwarz and E.  Witten, {\it Superstring Theory}
(Cambridge U.P., 1987).}

\lr \quev {C. Burgess and F. Quevedo,  \np B 421 (1994) 373. }
  \lr \nahm { W. Nahm, \np B124 (1977) 121. }
 \lr \kkkk {  E. Kiritsis and C. Kounnas,  ``Curved four-dimensional spacetime
as infrared regulator in superstring theories", hep-th/9410212. }
\lr \sen{A. Sen, \pr D32 (1985) 2102; \prl 55 (1985) 1846.}
\lr \hulw { C. M. Hull and E.  Witten,  \jnl \pl, B160, 398, 1985. }

\lr\hult { C. Hull and P. Townsend, \jnl \pl, B178, 187, 1986. }
\lr \gps {S.  Giddings, J. Polchinski and A. Strominger, \jnl  \pr,  D48,
 5784, 1993. }

\lr \ghr { S. Gates, C. Hull and M. Ro\v cek, \np B248 (1984) 15.}
\lr \jon {C. Johnson, \pr D50 (1994) 4032.}
\lr \landau {L.D.  Landau and E.M.  Lifshitz, {\it Quantum Mechanics} (Pergamon
Press, N.Y., 1977).  }
\lref \horts {G.T. Horowitz and A.A. Tseytlin, \pr  D51 (1995) 2896. }

%\lr \sfetsos { K. Sfetsos, \pl B324 (1994) 335. }
\lr \attick {J.J. Attick  and E. Witten, \np B310 (1988) 291. }
\lr\fund{A. Dabholkar, G. Gibbons, J. Harvey and F. Ruiz Ruiz, \jnl \np, B340,
33, 1990;
D. Garfinkle, \jnl \pr, D46, 4286, 1992; A. Sen, \jnl \np, B388, 457, 1992;
D. Waldram, \jnl \pr, D47, 2528, 1993. }

\lr\planetach{ J. Garriga and E. Verdaguer, Phys. Rev. {\bf D43} (1991) 391.}

\lr \duality { L. Brink, M. Green and J. Schwarz, \np B198 (1982) 474;
K. Kikkawa and M. Yamasaki, \pl B149 (1984) 357;
N. Sakai and I. Senda, { Progr. Theor. Phys. }  75 (1984) 692. }

\lr \canon {A. Giveon, E. Rabinovici and G. Veneziano, \np B332 (1989)167;
K. Meissner and G. Veneziano, \pl B267 (1991) 33;
E. \' Alvarez, L. \' Alvarez-Gaum\' e and Y. Lozano, \pl B336 (1994) 183. }

\lr \kkl {E. Kiritsis, C. Kounnas and D. L\" ust, \pl B331 (1994) 321.}
\lr \kk {E. Kiritsis and  C. Kounnas,  \pl B320 (1994) 361.}
\lref \kallosh { E. Bergshoeff, R. Kallosh and T. Ort\' \i n, \jnl \pr,  D47,
5444,
1993; E. Bergshoeff, I. Entrop and R. Kallosh,
\jnl \pr, D49, 6663, 1994.   }

%\lr\sfettt {K. Sfetsos, \prd    }
\lr\tseta{A.A. Tseytlin, \pl B317 (1993) 563;
K. Sfetsos and A.A. Tseytlin, \pr D49 (1994) 2933.  }

\lr\frats { E.S. Fradkin and A.A. Tseytlin , \pl B163 (1985) 123. }
\lr\tsey{  A.A. Tseytlin, \pl B202 (1988) 81.  }
\lr\tset { A.A. Tseytlin, \np B350 (1991) 395.}
\lr\metsa {R.R. Metsaev and A.A. Tseytlin, \np B298 (1988) 109. }
\lr\abo { A. Abouelsaood, C. Callan, C. Nappi and S. Yost, \np B280 (1987) 599.
}
\lr\bachas {C. Bachas and M. Porrati , \pl B296 (1992) 77.}
\lr\ferrara{S. Ferrara and M. Porrati, \mpl A8 (1993) 2497.}
\lr \sfets { K. Sfetsos and  A.A. Tseytlin, \np B427 (1994) 325.  }
\lr\rs {J.G. Russo and L. Susskind,   \np B437 (1995) 611.}
\lr\thermal {J.G. Russo, \pl B335 (1994) 168.}
\lr \napwi{ C. Nappi and E. Witten, \jnl \prl, 71, 3751, 1993.}

\lr\tsetl { A.A. Tseytlin, \pl B208 (1988) 221.}
\lr\polch{ J. Polchinski, Commun. Math. Phys. 104 (1986) 37.}
\lr\obrien{ O'Brien -Tan , McClain-Roth }
\lr \ginsp { W. Lerche, A. Schellekens and N. Warner, Phys. Repts. 177 (1989)
1; P. Ginsparg, in: {\it Fields, Strings and Critical Phenomena, } ed. by E.
Brezin and J. Zinn-Justin (Elsevier Science Publ., 1989).  }
\lr\salam{A. Salam and J. Strathdee, \np B90 (1975) 203.}
\lr\nielsen{ N.K. Nielsen and P. Olesen, \np B144 (1978) 376;
J. Ambjorn and P. Olesen, \np B315 (1989) 606; \np B 330 (1990) 193.}

\lr\anto{I. Antoniadis and N. Obers, \jnl \np, B423, 639, 1994}
\lr \nonsemi { D. Olive, E. Rabinovici and A. Schwimmer,  \jnl \pl, B321, 361,
1994.}
\lr\sfee{
 K. Sfetsos,  \jnl \pl, B324, 335, 1994; \pr D50 (1994) 2784.}
\lr\nons{A.A.  Kehagias and P.A.A.  Meesen, \pl B331 (1994) 77;
J.M. Figueroa-O'Farril and S. Stanciu, \pl B327 (1994) 40;
A. Kehagias, ``All WZW models in $D\leq 5$", hep-th/9406136. }

\lr \busc {T.H. Buscher, \pl  B194 (1987) 59; \pl B201 (1988) 466.}
 \lr\rocver{ M. Ro\v cek and E. Verlinde, \np B373 (1992) 630. }

\lr\rutse{J.G. Russo and A.A. Tseytlin, {`` Exactly solvable string models
of curved space-time backgrounds"}, CERN-TH/95-20, Imperial/TP/94-95/17,
hep-th/9502038.}
\lr\melvint{A.A. Tseytlin,  \pl B346 (1995) 55. }

\def\np {  Nucl. Phys. }
\def \pl { Phys. Lett. }
\def \mpl { Mod. Phys. Lett. }
\def \prl { Phys. Rev. Lett. }
\def \pr  { Phys. Rev. }

\def \ijmp {Int. J. Mod. Phys. }

\baselineskip8pt
\Title{\vbox
{\baselineskip 6pt{\hbox{CERN-TH.7494/94 }}{\hbox
{Imperial/TP/94-95/3 }}{\hbox{hep-th/9411099}} {\hbox{
revised
   }}} }
{\vbox{\centerline { Constant magnetic field in closed string theory:   }
\centerline {an exactly solvable model }
}}

\vskip -20 true pt

\centerline  { {J.G. Russo\footnote {$^*$} {e-mail address:
 jrusso@vxcern.cern.ch
} }}

 \smallskip \smallskip

\centerline{\it  Theory Division, CERN}
\smallskip

\centerline{\it  CH-1211  Geneva 23, Switzerland}

\medskip
\centerline {and}
\medskip
\centerline{   A.A. Tseytlin\footnote{$^{\star}$}{\baselineskip8pt
e-mail address: tseytlin@ic.ac.uk}\footnote{$^{\dagger}$}{\baselineskip8pt
On leave  from Lebedev  Physics
Institute, Moscow, Russia.} }

\smallskip\smallskip
\centerline {\it  Theoretical Physics Group, Blackett Laboratory}
\smallskip

\centerline {\it  Imperial College,  London SW7 2BZ, U.K. }
\bigskip
\centerline {\bf Abstract}
\medskip
\baselineskip8pt
\noindent
We consider a  simple model describing a closed bosonic string
in a constant magnetic field.  Exact conformal invariance
demands also the presence of a non-trivial metric and
antisymmetric tensor (induced by the magnetic field).
The model is invariant under target space duality in a compact
Kaluza-Klein direction  introduced  to couple the magnetic field.
Like open string theory in a constant gauge field,
or closed string theory on a torus, this model can be
straightforwardly quantized and solved  with its spectrum
of states and  partition function  explicitly computed.
Above some critical value of the magnetic field an infinite
number of states become tachyonic, suggesting a presence  of
phase transition. We also construct heterotic string
generalisations of this  bosonic model in which the constant
magnetic field is embedded either in the Kaluza-Klein or
internal gauge group sector.

\Date {November 1994}

%\draftmode
\noblackbox
\baselineskip 14pt plus 2pt minus 2pt
%\baselineskip 20pt plus 2pt minus 2pt
%%%%%%%%%%%%%%%%%%%%%%%%%%%%%%%%%%%%%%%%%%%%%%%%%%%%%%%%%%%%%%%%

\vfill\eject

%%%%%%%%%%%%%%%%%%%%%%%%%%%%%%%%%
\newsec{Introduction}
%%%%%%%%%%%%%%%%%%%%%%%%%%%%%%%%%%%%%

The study of simple exactly solvable models may provide a way to understand
vacuum properties in string theory
 and discover its hidden symmetries.
In the case of open string theory, the simplest
example of a non-trivial
background is  a  constant  abelian gauge field.
It represents a conformally invariant theory,  and various aspects of string
theory can be consistently analysed  with a  constant field strength
as  a free external parameter.
For example, one  can  easily find   the exact
tree-level (disc) expression for the partition function  $Z_0$  and  (since in
the open string theory the massless  effective action $S$ is
proportional to the renormalised value of $Z_0$
\refs{\frats,\tsey})  discover that  the $\del F$- independent  part of the
exact tree-level $S(F) $ is given by   the Born-Infeld action.
%\foot{This can be generalised to superstring case as well, see e.g. refs.
%\tset,\metsa }
One can also compute
% rederive this result using  beta-function approach \abo,
find  the 1-loop (annulus) correction to the partition function
\refs{\frats,\abo} and  study the open string   behaviour  in the constant
electric or magnetic fields
(e.g. analyse tachyonic instabilities and existence of a critical field  \burg,
string pair creation  by an  electric field \bachas,  phase transition in a
magnetic field \ferrara, dependence of the Hagedorn temperature on the electric
field \incer, etc.).

  In this open string  model  the target space metric is the Minkowski one,
i.e. the deformation of the geometry caused  by  the presence of the
electromagnetic
stress-energy  is ignored.
This   is a good approximation for weak  fields  much smaller than the Planck
scale ones.
It would be very interesting to
include  the  back reaction, i.e.  to treat   the  geometry and the
 gauge field
 on the same footing.
 Unfortunately, in the case of the open string theory a complete treatment
requires including  higher loop orders (to incorporate the bound-state closed
string sector) and is hardly  tractable.

Given that the physically relevant  string  models are expected to be
{\it  closed } string ones
\refs{\green,\gro}
a natural  question is whether there exists a simple solvable analogue
of the   constant abelian gauge field problem
 in the closed bosonic, superstring and heterotic string theories.
It is known how   to incorporate some simple types of external
 parameters  (e.g.  constant  moduli   of a target space torus
or a temperature) in closed string theory.
Here we  suggest how to include another one,  representing the constant
magnetic field strength.

 One may wonder how important is  actually the  influence
of the magnetic field on the geometry. In principle, the new scale $f\inv$
introduced by  the magnetic field (${\cal H}= R\inv f$)
and can be arbitrary.   In practice,  to study
non-trivial stringy effects one is to consider $f\inv$  to be
of a Planck scale.
 In fact, the present model is  a ``conformal improvement"
of  the one investigated in \refs{\rs,\thermal}  where the change in the
geometry due to the presence of the magnetic
field was ignored.
 In ref. \thermal \
 it was observed that the corresponding
thermodynamic partition function of a  string gas in a magnetic
field  diverges if $f$ exceeds some critical value. It was suggested  that a
phenomenon  analogous to
 the Meissner effect in  type-I superconductors (leading to a critical curve on
the magnetic  field -- temperature  diagram), where strong fields restore the
hidden symmetry
of the
 theory,   may  take place in string theory.
This was supported by
the appearance of  infinitely many extra massless states at a  critical value
of the field (the emergence of tachyons at strong magnetic
fields was found also in the open string model  \ferrara).
 The critical magnetic field has  Planck value ($f \sim 1/\sqrt{\a'}$), i.e.
 it produces a
 a strong deformation of the geometry, with  the curvature being proportional
to $f^2$.
That means the back reaction cannot be ignored,  and
to address these important issues one is to  repeat the analysis  starting
with a consistent conformally invariant model.

To couple an external magnetic field  (described  by a vector potential
$A_i$) to a closed bosonic model one may follow the Kaluza-Klein idea,
introducing a periodic  ``internal" coordinate $\p$ (compactified on a circle
of radius $R$)  and adding the terms $ (\del \p )^2$ and $  A_i\del x^i \del\p$
to the free string Lagrangian (in the heterotic string case the role of $\p$
may be played by a Weyl fermion of an internal gauge sector).
The resulting \sm  is not, however, conformally invariant. This is not
surprising since the energy of the magnetic field should modify
the space-time geometry in the non-compact $(t,x^i)$ directions. To find a
conformally invariant model, one should start with the  full string
(Einstein-Maxwell-type) equations, plug in the ansatz for the magnetic field,
determine the corresponding metric, antisymmetric tensor and dilaton, then go
to the next order in $\a'$, etc.

Remarkably, there exists a simple conformal  model \horts\
which  gives an  exact solution to  this seemingly
complicated back reaction problem.
It has  a curved  metric
 parametrised by  $A_i=-\ha F_{ij} x^j $  (which is nothing but the metric of
the product of the group space of the Heisenberg group and a line)
and the antisymmetric tensor  proportional to $F_{ij}$
(and constant  dilaton). Introducing the ``light-cone" coordinates
$v= \p +  t, \ u = \p-t$
one learns that the corresponding string Lagrangian
%\eqn\sss{ L= \del u \bd v + 2 A_i (x) \bd x^i \del u + \del x^i \bd x_i \ , }
is formally  ($\p$ is periodic!)
of a ``plane- fronted wave" type, i.e. has a null covariantly constant Killing
vector (see e.g. \refs{\plane,\tsenul,\duval,\kallosh})
and,   provided the Maxwell equation $\del_i F^{ij}=0$ is satisfied,
represents a string theory solution to all orders in $\a'$  \horts.
In the limit of non-compact $\p$ ($R\to \infty$) in addition to having  ``plane
wave" interpretation,   this model is equivalent \horts\
to  a  WZW model based on a non-semisimple group
(with the basic  $D=4$ example being the model of ref. \napwi, see also
\refs{\kk,\nonsemi,\sfee,\anto,\sfets,\nons}).

The purpose  of the present paper is to investigate
the properties of this  model   by solving it explicitly, i.e.
computing the spectrum of  states  and  the partition function.
This model is easier to solve  than, e.g.,  WZW models based on simple groups;
in particular, the possibility of fixing the light-cone gauge demonstrates its
unitarity.
  We shall
confirm  the presence
of the  tachyon instabilities and  a phase transition for a critical value of
the magnetic field.   We expect  that  it should be possible
to   determine the scattering amplitudes
and thus  an
effective action for  extra ``massless" fields that emerge at the critical
magnetic  field. That may help   to unravel new string symmetries.

The essential  part of  the definition of our model is that $\p$ is a
  ``Kaluza-Klein" coordinate,
i.e. is   compactified on a circle of radius $R$.
Interestingly, the target space duality symmetry \duality\
interchanging momentum
and winding modes and $R$ with $\a'/R$  which is present in  the free  theory
remains an   exact symmetry of  this  model (in particular, of its  spectrum
and the  partition function)  for  arbitrary  magnetic field.
Another  important property is that  while
 in Kaluza-Klein  theory the role of charge is played by the linear  momentum
in a compact direction, here the ``charge"
(to which the magnetic field  is coupled in the mass formula)
is represented   by  the sum  of  the ``left" string momentum  $Q$ (equal to
the duality
invariant combination   of the linear  and winding momenta)
and the energy $E$ of the string. The appearance in the string mass formula
 of the term $Ef$ (in addition to $Qf$) which is  linear  in the energy
is a direct consequence of the non-flat geometry of our background (or,
equivalently,  of  its conformal invariance).

The plan of this  paper is the  following. In Section  2 we describe some
general aspects of the model, including its geometrical
interpretation,   relation to non-semisimple WZW models, target space
 duality, point-particle limit and  explicit solution of  the tachyon equation.

In Section 3 we solve the classical  string equations,
express the solution in terms of the free fields,
perform  the complete   quantisation of  the theory in the ``light-cone" gauge
and analyse the resulting spectrum of states.

Section  4 describes  the calculation of the partition function on the torus,
first  starting from  the path integral  and then using the operator approach
(with equivalent results). We discuss the new  singularities in  the partition
function related to the presence of  tachyon  instabilities in the spectrum.

In Section 5 we discuss the  supersymmetric extensions of the bosonic model.
In particular, we construct the exact conformal heterotic string models
in which the constant magnetic field is embedded either in the bosonic
(Kaluza-Klein) or in the fermionic (internal gauge group)  sector.
 Some conclusions and open problems are presented in  Section 6.

%%%%%%%%%%%%%%%%%%%%%%%%%%%%%%%%%%%%%%%%%
\newsec{Description of the model}
%%%%%%%%%%%%%%%%%%%%%%%%%%%%%%%%%%%%%%%%%%%%%
\subsec{ Abelian vector coupling in closed string theory }
%%%%%%%%%%%%%%%%%%%%%%%%%%%%%%%%%%%%%%%%%%%%%%%%
To motivate the discussion of  our specific string model
let us consider the following problem: how
  to ``embed"  a spatial (or magnetic)  part of
Maxwell equations in  {\it closed}
string theory?
A simple application of the Kaluza-Klein idea leads to the following model (see
e.g.   \rs )
\eqn\actio{ I={1\over 4\pi\alpha '}\int d^2 \s\big[ \sqrt{g} \big( -
\del_\alpha t  \del^\a t  + \del_\alpha x^i \del^\a x_i
+\del_\alpha \p \del^\a \p} $$  +2A_i(x)\del_\alpha x^i \del^\a \p \big) +
2iA_i(x)\epsilon^{\alpha\beta}\del_\alpha x^i \del_\beta \p \big]\ ,
$$
where $x^\m= (t, x^i)$ are coordinates of a flat space-time  ($i=1, ..., D-2$)
and $\p $
 is a compact ``internal" coordinate  used to couple the gauge field to the
string. The vector and axial gauge fields  corresponding to the ``off-diagonal"
 $(x^\m\p)$-components of the metric and the antisymmetric tensor field  are
taken to be equal. As a result, only the ``right" part  of $x^i$ is coupled to
$A_i$, i.e. has a gauge charge.\foot{
 In this sense such model is  a bosonic analogue of the
heterotic string. Assuming that $\p$ can be fermionised
(and adding extra free degrees of freedom) one may  also  directly
embed this model into the heterotic string theory (see Section 5).}
% \foot{ The extension of the present results to the heterotic string theory
%%%will %be carried out in Section 7.}
In the conformal gauge, the Lagrangian density corresponding to action \actio\
becomes  ($I\equiv  {1\over \pi\alpha'}\int
d^2\sigma  L$)\foot{Here we consider the Euclidean world sheet and use the
following notation: $\del= \ha (\del_1 - i \del_2), \ \bd = \ha (\del_1 + i
\del_2), \
z= \s_1 + i \s_2, \ \bar z = \s_1 -i\s_2, \ d^2 \s \equiv d\s_1 d\s_2. $
In the case of the Minkowski world sheet used in Section 3 $\del$ and $\bd$
should be replaced by $\del_+ = \ha (\del_\t + \del_\s), \ $ and
$ \del_- =
\ha (\del_\t - \del_\s)$,  and $z, \bar z$ by $\s_\pm = \t\pm \s$,
and the overall sign of the action should be reversed. }
\eqn\lagrone{ L= - \del t \bd t +   \del \p \bd \p
+ 2A_i(x) \bd x^i \del \p + \del x^i \bd x^i \ .}
Eq. \lagrone\ does not, however,  represent a  conformally invariant model,
i.e. such vector coupling   does not  satisfy the  string equations,
implying that one needs to  add   also other background fields.

 It is remarkable that a  conformal  model can be  obtained by  a  simple
modification of \lagrone, introducing  non-trivial
 metric and  the antisymmetric tensor  components in the ``$i0$" direction,
\eqn\acti{ I={1\over 4\pi\alpha'}\int
d^2 \s \big[ \sqrt{g}  \big( - \del_\a t \del^\a t  + \del_\alpha x^i \del^\a
x_i
+\del_\alpha \p \del^\a \p } $$
+2A_i(x)\del_\alpha x^i \del_\beta ( \p -t) \big)
+2iA_i(x)\epsilon^{\alpha\beta}\del_\alpha x^i \del_\beta (\p -t)\big] \ .
$$
As a result, the $x^\m$-part of the model is no longer flat, with non-trivial
metric and antisymmetric tensor ``compensating" for the distortion of the
geometry by the magnetic field.
The conformal gauge Lagrangian corresponding to eq. \acti\   takes  the
form
\eqn\la{
 L =-\del t \bd t +   \del \p \bd \p
+ 2A_i(x) \bd x^i (\del \p-\del t)  + \del x^i \bd x^i \ ,
}
i.e. (up to  a total derivative)
\eqn\lag{
 L= \del u \bd v  + 2A_i(x) \bd x^i \del u + \del x^i \bd x^i\ ,}
$$  \ \ u\equiv  \p -t \ , \ \ \ \ \  v\equiv   \p  + t \ .
$$
The  action is invariant under a  gauge transformation of $A_i$
combined with a coordinate transformation  of $v$
%(since $u$ is not transforming $\a$ must be periodic to preserve periodicity
%of $\p$)
\eqn\gau{  \ A_i'=A_i  + \del_i \a \  , \ \ \ \  v'=v - 2\a \ , \ \ \ \a=\a(x)\
.  }
This model  belongs to the general class of models considered in  \horts\
(see also \tsenul)  and represents an {\it exact} conformal field theory (to
all orders in $\a'$) provided  $A_i$ satisfies  the magnetic part of the
Maxwell's equations in flat space\foot{One may also consider the model \horts\
 $\ L= \del u' \bd v'  + k \del u'\bd u' + 2A_i(x) \bd x^i \del u' + \del x^i
\bd x_i
\ , \ \ k= \const\ . $
The two models are formally related
($v= v' + ku', \ u=u'$) as in   the case of non-compact  coordinates (the
dependence on $k$ can be absorbed into a rescaling of $\a'$ in the physical
quantities like partition function).}
\eqn\maxw{
\del_i F^{ij}=0\ , \ \ \ \ \ F_{ij} \equiv \del_i A_j - \del_j A_i \ .
}
In the main part of this paper we will study  a particular model
with $A_i=-\ha f\epsilon_{ij} x^j,  \  (i,j=1,2)$, representing a constant
magnetic field
in 3-space.\foot{A simple reason why
  the corresponding \sm \lag\ is conformal
is the  following (for a rigorous argument see \horts). If one first does not
introduce a source coupled
to $v$,  the path integral over $v$ imposes a constraint on $u$
and thus gives a free theory for $x^i$. It remains then to check
conformal invariance in the $uu$ and $ui$ directions.
The corresponding $\beta$-functions can not contain $\a'$-corrections
since $L/\a'$  (with $A_i$ quadratic in $x^i$) is invariant
under the simultaneous  rescaling   $v\to \l^2 v,\   x^i\to \l x^i$ and $\a'\to
\l^2 \a'$.
Thus  $\a'$ can be given an arbitrary value
by a coordinate transformation but the $\beta$-functions
are assumed to be covariant tensors transforming
under the above rescalings in the standard  way (notice that $u$ is not
rescaled).
To see  why the  1-loop conformal invariance
conditions are satisfied one may use the observation that this model
is formally (ignoring the issue of  periodicities of coordinates)  dual
to a  flat space model, see Section 2.4. }

 Since $A_i$ is   dimensionless
like a  component  of the  metric,   $f$ has an
inverse length dimension. The ``physical" (dimension cm$^{-1}$) vector field is
obtained by dividing $A_i$ over
 the radius $R$ of the compact dimension $\phi$, i.e.
 the proper   magnetic field  is $ {\cal H} = fR\inv$
(for simplicity  in what follows we shall still call $f$ the magnetic field).
 $f\inv$ defines another scale of our model which is independent of
$R$ and $\a'$ (in the case of a heterotic model  with  the magnetic field  in
the internal gauge sector the scale  $R$ will be absent, or, effectively,   $R=
\sqrt {\a'/2}$, see Section 5).
Since  the curvature  and the antisymmetric tensor field strengths
corresponding to our model satisfy $R\sim H^2\sim F^2 \sim f^2$, by a choice of
$f$
their scale can be made arbitrarily larger than the Planck scale
($\sqrt{\a'}$).

The  $D$-dimensional background  corresponding to \lag\ is a mixture of
a curved metric $G_{MN}$  and the antisymmetric tensor $B_{MN}$
 ($x^M= ( t, x^i, \p)$).
The metric
\eqn\metricone{
 ds^2= G_{MN} dx^M dx^N= dudv + 2A_i(x) dx^idu + dx^idx_i \
 }
is of a plane-fronted wave type, i.e. has one covariantly constant Killing
vector.\foot{ The inverse metric is
$
 G^{uv}= 2    ,  \  G^{ui}=  G^{uu } = 0  , \
G^{vi} = - 2 A^i  ,    \  G^{vv}=4A_iA^i    ,
 \  G^{ij} =\delta^{ij}   ,  \ \det G\inv =-4  .
$}
The  generalised connection
which enters the string propagation equation
\eqn\conn{ \del\bd x^L + \hG^L_{-MN} (x) \del x^M \bd x^N =0\   , \ \ }
\eqn\con{ \hG^L_{\pm MN}  =\G^L_{MN} \pm
 \ha {H^L}_{MN} \ ,  \ \ H_{LMN} = 3\del_{[L} B_{MN]} = \{ H_{uij} = F_{ij},\
0\}\ ,  }
has the following non-zero components:
 $
\hat\Gamma^i_{-uj} =  -   {{ F^i}_j} \   ,   \  \
\hat\Gamma^v_{-ij } = 2 \del_i A_j \  ,  \  \
\hat\Gamma^v_{-ui} = - 2 A^j  { F}_{ij}\ . $
The only non-trivial components of the corresponding generalised curvature
tensor and Ricci tensor are\foot{The standard  curvature  of the metric
contains extra
$\del F $ and $F^2$-terms.}
 \eqn\ricciten{ {\hat R}^v_{- ijk }= -2  {\hat R}^i_{- u jk } =
2\del_i { F}_{jk}  \ ,  \ \ \ \
\hat R_{-ui}= -\del_j {{{ F}^j}_i}\ .
}
If $A_i$ satisfies \maxw\
the space is thus  Ricci-flat  in generalised sense
($\hat R_{MN}=0$, i.e. $ R_{MN} = \fourth H_{MKL} H_{N}^{\ KL}, \   \nabla_K
H^{MNK}=0 $).
 Moreover,  if $A_i$ has a constant field strength,  it is parallelisable,
i.e. $ {\hat R}^M_{\ NKL} =0$.\foot{It should be noted that the
parallelisability is a natural  string theory generalisation of the flatness
property
in the point-particle theory case.} The  resulting space is,
in fact,  a  group space of a non-semisimple group, see  \horts\ and below.
Introducing a vierbein and  computing the  components of the Lorentz connection
one finds that for $F_{ij}=\const.$,   $de^m + \omega^m_{\ n} e^n=0$
can be identified with the Maurer-Cartan equation  from which
one can read off the structure constants (proportional to $F_{ij}$) of the
corresponding non-semisimple algebra.

%%%%%%%%%%%%%%%%%%%%%%%%%%%%%%%%
%%%%%%%%%%%%%%%%%%%%%%%%%%%%%%%%%%%%%%%%%
\subsec{Constant field strength case  and relation to non-semisimple WZW
models}
%%%%%%%%%%%%%%%%%%%%%%%%%%%%%%%%%%%%%%%%%%%%

In what follows we shall specialise to  the case of a constant magnetic  field
strength
\eqn\parr{   \ A_i=-\ha { F}_{ij} x^j \ ,  \ \ \ F_{ij} = \const \ , \ \ \ \
i,j= 1, ...., D-2\ .  }
When $u,v$ are {\it non-compact},  the   spaces corresponding to \lag,\parr\
 can be interpreted \horts\ as ``boosted" products of group spaces,
or, equivalently, as  group spaces corresponding to
 non-semisimple groups \napwi\ (see also \refs{\nonsemi,\sfee}).
This can be demonstrated by
 first putting ${ F}_{ij}$  (by a  rotation of  $x^i$) into  a
 block-diagonal
form, so that its elements are represented by constants
$f_1, ..., f_{[D/2-1]}$. The  Lagrangian
 is  then (we split $x^i$ into pairs representing 2-planes)
\eqn\jkl {
 L= \del u \bd v  + F_{ij} x^i \bd x^j \del u + \del x^i \bd x^i }
$$ =
\del u \bd v + \sum_{s=1}^{[D/2-1]}
\big(f_s \ep_{i_s j_s } x^{i_s} \del u \bd x^{j_s } +
\del x^{i_s}  \bd x_{i_s} \big) \ . $$
 The first non-trivial
case is that of $D=4$, i.e. ${ F}_{ij} = f \ep_{ij}$.
The  corresponding model
($x_1=r \cos \theta, \ x_2 = r \sin \theta $)
\eqn\moo { L=\del u \bd v +  f \ep_{ij} x^i \bd x^j \del u
+ \del x^i\bd x_i }
$$=\del u \bd v + f  r^2 \bd \theta \del u
+ \del r \bd r + r^2 \del \theta \bd \theta\ , $$
is equivalent to the $E^c_2$ WZW model of ref.\napwi.
 In fact, if  $\p= \ha (u +v)$ is   assumed to be  {\it non-compact}, then
 $f$ can be set equal, e.g.,  to   --1 by a rescaling of
$u,v$.
The coordinate  transformation
%(defined  when $\p$ is non-compact)
 \eqn\coo{x_1=y_1 + y_2 \cos u\    , \ \ \
\ x_2 = y_2 \sin u\ , \ \ \ \ v= v' + y_1 y_2  \sin u \ , }
  puts \moo\
in the form (up to a total derivative term)\foot{The  Lagrangian \moo\ can be
put also in the following equivalent form
(see e.g. \anto)
$\  L = \del u \bd v   + f x^2  \del u \bd u  +
 f \ep_{ij} x^i(\del u \bd x^j - \del x^j \bd u)
+  \del x_i \bd x^i     .  $     }
\eqn\fgf{ L =\del u \bd v' + \del y_1 \bd y_1 + \del y_2 \bd y_2
+ \ 2 \cos u\ \del y_2 \bd y_1  }
\eqn\fgff{\equiv  \del u \bd v' +  (g_{ij} + b_{ij})(u) \del y^i \bd y^j
\ , }
which is obtained from the  $R\times SU(2)$ WZW action
by a singular boost
and rescaling of the level $k$ or $\a'$
 (see \refs{\sfee, \sfets}).

All higher $D$ models are related \horts\  to similar WZW models
based on direct products of $SL(2,R)_{-k}, \ SU(2)_k$ and $R$ factors,
 or, equivalently, on corresponding non-semisimple groups.
The parameters $f_s$  are  essentially
the rescaled levels $k_n$ of the factors.

In the rest of this paper we shall consider the simplest  $D=4$
model \moo; however,
all our results (exact solution of the string  equations, light-cone
quantisation, spectrum, partition function) can be straightforwardly
generalised to
the case of \jkl\  with  an arbitrary $D$.

In order to describe  the
interaction  of a closed string with a  magnetic field
we have assumed that  $\p$ is  compactified on a circle
 with period $2\pi R$.
In the limit $R\to \infty $ our model is equivalent   to the WZW
model of \napwi.
 Though   the  construction of \napwi\
of  a non-semisimple  WZW model formally goes through also in the case of a
periodic  $\p=\ha (u+v)$
%\foot{The corresponding algebra is of course the same
%as used in \napwi; it is not clear, however, if the model
%with periodic $\phi$ can be obtained from a  $R\times G$ WZW model (where $G$
%%is %a semisimple dimension 3 group)  by taking a singular limit as described
%%in
%\sfts.}
we  shall not use the current algebra  relations
in the explicit solution of  our  model  in Sections 3 and 4.\foot{
 The  current algebra  approach to the solution of the model of \napwi\ in the
case
of a non-compact $\p$ was developed  in  \refs{\kk,\kkl}. }

%%%%%%%%%%%%%%%%%%%%%%%%%%%%%%%%%%%%
\subsec{ Dimensional reduction interpretation }
%%%%%%%%%%%%%%%%%%%%%%%%%%%%%%%%%%%%%%%%%%%%%%
In trying  to couple an abelian gauge field to a closed string theory in a
conformally invariant way we have suggested
the simple  model \lag. In order to further clarify the  ``magnetic field"
interpretation of  $A_i$ let us  reinterpret this model in the Kaluza-Klein
fashion, assuming that  $\p$ is a compact internal dimension
(see also  the discussion in \horts).

One can rewrite the Lagrangian  \lag\ in the form
 \eqn\lagr{ L= -  [ \del t  + A_i (x) \del x^i][\bd t  +  A_i (x) \bd x^i]
 +   \del x^i \bd x_i +  A_i(x)  (\del x^i \bd t - \bd x^i \del t) \ }
$$ + [ \del \p  + A_i (x) \del x^i][\bd \p +   A_i (x) \bd x^i]
  - A_i (x) (\del x^i \bd \p - \bd x^i \del \p ) \ . $$
For definiteness, we shall assume that  we start with the $D=5$ model and
represent it as  $D=4$ model with extra vector couplings.
Interpreting $\p$ as a  ``fifth" internal coordinate we find the corresponding
four dimensional background
\eqn\did{ ds^2_4= -  [ dt +  A_i(x)  dx^i]^2 + dx_idx^i \ , \ \ \ \  \ \
B_{it}= A_i (x) \ , }
$$
 \      {\cal A}_i  = - {\cal B}_i=   A_i(x) \ . $$
In addition to a curved metric and antisymmetric tensor backgrounds
there are  also vector $ {\cal A}_i  $ and axial-vector $ {\cal B}_i$ fields
equal (up to sign)  to $A_i$.
When $A_i$ satisfies \maxw\
 this is an exact (all-order in $\a'$)  solution corresponding to  the $D=5$
bosonic string
 effective action for $G_{MN}, B_{MN}$ and dilaton
$\Phi$ dimensionally reduced
to $D=4$
\eqn\acttp{  S_4 = \int d^4 x \sqrt {\hat G }\  e^{-2\Phi + \s}    \ \big[
  \   \hat R \ + 4 (\del_\m \Phi )^2 - 4 \del_\m \Phi \del^\m \s }
$$  - {1\ov 12} (\hat H_{\m\n\l})^2\  - \fourth e^{2\s} ({ \cal F}_{\m\n}
({\cal A}))^2
-\fourth  e^{- 2\s} ({\cal F} _{\m\n} ({\cal B}))^2
  + O(\a')   \big]  \  , $$
where  we have defined $G_{55}\equiv  e^{2\s}\ $ and
  $$   {\cal F}_{\m\n} ({\cal A}) = 2\del_{[\m}
{\cal A} _{\n]}  \ ,   \ \ {\cal F} _{\m\n}({\cal B}) = 2 \del_{[\m} {\cal B}
_{\n]}  \  ,
\ \ \  {\cal A} _\m\equiv   G^{55}  G_{\m 5}\ ,  \ \ {\cal B} _\m \equiv
B_{\m 5}\ , $$
\eqn\fgfg{\hat G_{\m\n} \equiv  G_{\m\n} - G_{55}{\cal A}_\m {\cal A}_\n
\ , \ \ \
\hat H_{\l\m\n} = 3\del_{[\l} B_{\m\n]} - 3 {\cal A}_{[\l} {\cal F}_{\m\n]}
({\cal B})
\ .  }
Since   ${\cal A}_i$ and  $-{\cal B}_i$  are equal on  our solution
 (this is directly related to the fact that the scalars $\Phi$ and $\s$ are
constant), they can be treated as a single vector field. In fact,  assuming
that $\Phi=\const $,  $\ \s=0$   and introducing the  two vectors
$V_\m = \ha ( {\cal A}_\m -{\cal B}_\m) $  and $ W_\m =\ha ({\cal A}_\m + {\cal
B}_i) $  we get the effective action
\eqn\acttpp{  S_4 = k_0 \int d^4 x \sqrt {\hat G }\ \big[
  \   \hat R \   - {1\ov 12} (\hat H_{\m\n\l})^2\  - {1\ov 2}  ({ \cal
F}_{\m\n} (V) )^2
- {1\ov 2}   ({\cal F} _{\m\n}(W) )^2
  + O(\a')   \big]  \  . }
We conclude that the  solution  \did\
describes a  curved space-time with the  antisymmetric tensor $\hat H_{tij}=
H_{tij}= F_{ij}$  and one non-vanishing vector field   $\ V_\m =(0, A_i)$ with
the magnetic field strength  $F_{ij}$.

Introducing
the vierbein basis $e^m, \ m=(\hat 0, \hat i)$ for   the metric \did\
\eqn\vier{ e^{\hat 0}= dt + A_i (x)  dx^i\  , \  \ \ \  e^{\hat i}=dx^i \  ,
}
one finds that  the components of the
Lorentz connection are
\eqn\lore{ \omega^{\hat 0}_{\  \hat i} = \ha F_{ij}  e^{\hat j}\ , \ \ \ \
\omega_{\hat i \hat j} = \ha   F_{ij}  e^{\hat 0}\  .   }
When  $F_{ij}= \const $
we thus get the Maurer-Cartan equation:
 $ \ \ de^{l}= - \ha  C^l_{\ mn} e^m \wedge e^n$, $ \  C^{\hat 0}_{\ \hat i\hat
j} = -F_{ij}, \ C^{\hat k  }_{\ mn} =0$. The  dual  basis of the vector fields
thus satisfies   the Heisenberg algebra (it can be put into a canonical form by
``diagonalising" $F_{ij}$ as in \jkl)
\eqn\heis { [E_i,E_j] = -F_{ij} E_0\ , \ \ \ \  [E_i, E_0]=0 \ ,  \ \  \
F_{ij}=\const \ . }
The metric \did\  thus represents the group space of the Heisenberg group (or
its product with $R$ in the case when $D-1$ is odd).\foot{We are grateful to G.
Gibbons for this remark.}
This of course is directly related to the fact that
the higher dimensional  metric \metricone\  of which \did\ is a dimensional
reduction  represents a group space of
a non-semisimple group.

The magnetic field is constant in the preferred frame  chosen in \did\
in which the metric  is stationary. As follows from \lore\  the
field strength ${\cal F}_{\m\n}(V) = (0, F_{ij})$  is actually covariantly
constant ($F_{\hat t \hat i}=0, \ F_{\hat i \hat j}=F_{ij}$,
$\ \omega_{\hat i \hat k} F_{\hat k \hat j} +
\omega_{\hat j \hat k} F_{\hat i \hat k}=0$).
 The metric \did\ can be arbitrarily close
to the  flat one depending on the strength of the magnetic field.
It is not asymptotically flat ($R\sim F^2$)
 for non-zero $F_{ij}$
(i.e. it is  supposed to describe only the region of space-time
where the magnetic field is uniform).

One may wonder if there are similar uniform  magnetic field
 solutions of \acttp\
without antisymmetric
tensor field. The answer is no if  one does not introduce
 a non-trivial dilaton background.
The reason is that in addition to  the Eistein-type equation,  one is to
satisfy the dilaton equation that follows from \acttp.\foot{There exists the
axially symmetric static  leading-order solution \gibma\ which  is the
dilatonic generalisation of the Melvin solution. In  this ``flux-tube" type
solution
there is no antisymmetric tensor background but the magnetic field and dilaton
decrease  in the transverse directions [55]. }
This is impossible in terms of
  metric and vector field backgrounds only.
%One particular consequence of this observation is that the corresponding
%conformal \sm\  necessarily  contains  a term linear in the magnetic field:
%if only the metric were non-trivial, one could eliminate
%the linear term (assuming the expansion near flat space) by a change of the
%frame (cf. \rs).

To give an explicit example of the  solution \did\
let us  consider  with the $D=5$  model
($x^M=(t, \p, x^i, x^3),   \  i=1,2$) with  $A_i= -\ha f \ep_{ij} x^j, \ A_3=0$
representing a generic constant magnetic field in 3-space.
Then the $D=4$
metric
for the uniform magnetic field solution ($V_i= A_i$) is simply (cf. \moo)
\eqn\rotma{  ds^2_4 =  -(dt + \ha  f
  r^2 d\theta)^2 + dr^2
	     + r^2 d\theta^2 + dx^2_3  \ ,  }
 i.e. is the direct product of a $D=3$
space-time (group space of the 3-dimensional Heisenberg group) and  a line and
describes  a ``rotating universe" \horts.\foot{A
different  reduction
of the model   of \napwi\  giving   $D=3$   plane-wave type background
was considered in \sfets.}
  The non-trivial $D=3$ part of the metric \rotma\ satisfies
\eqn\eee{\ R_{\m\n} = \fourth H_{\m\l\r} H_\n^{\ \l \r} + {\cal F}_{\m\l}(V)
{\cal F}_{\n}^{\ \l}(V)  \ . }
  In the absence of the  gauge field  contribution the only solution in $D=3$
is   the anti de Sitter space (and its  analytic continuation or topological
 modifications). In fact, in $D=3$
$\ H_{\m\l\r}\sim \ep_{\m\l\r}$, $H_{\m\l\r} H_\n^{\ \l \r}\sim  G_{\m\n}$.
However, ${\cal F}_{\m\l} {\cal F}_{\n}^{\ \l}$  and thus $ R_{\m\n}$
are  no longer proportional to $G_{\m\n}$.
In the vierbein basis
$e^{\hat 0}= dt +    \ha  f r^2 d\theta, \  e^{\hat 1}= dr, \  e^{\hat 2} = r
d\theta, \ e^{\hat 3}=dx^3, $
one finds the following  non-vanishing vierbein components of the Riemann and
Ricci tensors corresponding to the homogeneous Heisenberg group space:
\eqn\rir{R_{\hat 0 \hat 1\hat 0 \hat 1}=R_{\hat 0\hat 2\hat 0\hat 2}= {1\ov 3}
R_{\hat 1\hat 2\hat 1\hat 2} = \four f^2 , \ \
R_{\hat 0\hat 0}=R_{\hat 1\hat 1}=  R_{\hat 2\hat 2} =  \ha f^2, \ \
R_{\m\n}R^{\m\n} = {3\ov 4} f^4 .  }
In  Sections 3,4  we shall analyse this solution  starting  directly with the
original $D=5$, or, equivalently (disregarding the  trivial  $x^3$-dimension)
$D=4$ model \lag\ itself,
and not with the
``dimensionally  reduced"  background \did.

%%%%%%%%%%%%%%%%%%%%%%%%%%%%%%%%%
\subsec{Target space duality invariance }
%%%%%%%%%%%%%%%%%%%%%%%%%%%%%%%%%%%%%
The spectrum and the  $S$-matrix  of the bosonic string theory in flat space
with one compact
dimension $\p$  are invariant under
 the duality  symmetry which  interchanges the
winding   and momentum modes corresponding to $\p$ and, at the same time,
inverts  the radius,  $R\to \a'R\inv $.  It is interesting that
 the inclusion of the gauge field  coupling
according to \acti\ or \lag\ preserves
the duality invariance. We shall now show that the model
\lag\ is ``self-dual", i.e. is invariant under the \sm duality
\busc\  with $A_i$ not transforming (see also \horts).\foot{The \sm duality
invariance  property is  true also for the dimensionally reduced
background \did\ with respect to duality in $t$ direction.}
 We shall also  see  the duality symmetry explicitly   in the spectrum
of our model derived in Section 3, and in the partition function computed in
Section 4.
\def \tR {\tilde R}

Starting with \la\  and gauging the isometry in the $\p$ direction by
introducing the 2d
gauge field $B, \B$ \rocver\ we get
\eqn\lagrsix {
 L=  R^2 (\del \vp + B)( \bd \vp  + \B)  - \del t \bd t
 + 2A_i(x) \bd x^i [ R( \del \vp  + B ) -\del t]
 }
$$
 + \del x^i \bd x^i + \a' (B \bd \td\vp - \B\del \td \vp) \ ,
$$
where  $\p= R\vp$, $ 0< \vp \leq 2\pi$ and  $\td\vp$ is the Lagrange multiplier
that sets the field
strength of $B,\B$ to zero.
Fixing the gauge $\vp=0$ and integrating over
$B,\B$ we get the    dual Lagrangian
\eqn\lagrseven {
\tilde L =  \tR^{2} \del \td \vp \bd \td\vp   - \del t \bd t
 + 2A_i(x) \bd x^i (\tR  \del \td \vp   -\del t)  + \del x^i \bd x^i \
}
$$=\del \td u \bd \td v  + 2A_i(x) \bd x^i \del \td u + \del x^i \bd x^i\ ,  $$
where $ \tR \equiv \a'/R , \   \td u = \tR \td \vp - t , \   \td v = \tR \td
\vp + t .$
Thus the theory  \lag\ is  invariant under $\vp\to \td \vp $, $R\to \tR $
with fixed  $A_i$.\foot{
It is possible to describe the theory using
the  manifestly duality invariant ``doubled" action \tset\
in which  both  the  original and dual fields $\vp$  and $\td\vp$ are present.
Starting  with \lagrsix,
 fixing  the  non-Lorentz-covariant gauge   $B_1=0$
 and integrating  over the $B_0$ component
($B\to  B_1-B_0,\   \B\to  B_1 +B_0$) one finds
$\
 L_{doubl.} =- \ha \dot \vp \td \vp ' +  \four R^2 \vp '^2 +
\four \tR^2 \td \vp '^2 + A_- ( R\vp ' + \tR \td \vp ')
 + A_-A_-  + O(t,x^i)  , $
% - 2 A_- \del_+ t  - \del_- t \del_+ t + \del_- x^i \del_+ x_i \ , $
where  prime is $\del_\s$,  dot is  $\del_\tau $ and
$ A_- = A_i(x)  \del_- x^i .$
$L_{doubl.}$ can be interpreted as a  Lagrangian for $\vp$  written in terms
of the coordinates and momenta  (represented by $ \td \vp '$), so that
 its
$\dot \vp$-independent part is proportional to the Hamiltonian.
It is  obviously symmetric under $\vp \to \td \vp $, $R\to \tR=\a'/R$.
 If one integrates out $\td \vp '$ (or $\vp$) one  gets back  to the original
\lag\ (or dual \lagrseven) action (for related canonical approach to duality
see \canon).}

\lr \klt{ C. Klim\v c\'\i k and A.A. Tseytlin, \pl B323 (1994) 305.}

The $F_{ij}=\const$  model \jkl\   has also rotational symmetry
in each of the 2-planes. For example, in the $D=4$ model \moo\ we
can thus make duality rotation in the $\theta$ direction.
Gauging $\theta\to \theta + a $  symmetry   get the following model
\eqn\modd{ L= \del u \bd v + \del r \bd r + r^2 (\del \theta + B)
(\bd \theta + \bar B)  +
f r^2 \del u (\bd \theta  + \bar B) +\a'( B  \bar\del  \tilde\theta -
\bar B \del \tilde \theta) \ , }
where $\tilde \theta$ is the Lagrange multiplier (dual coordinate) and $B, \bar
B$ are components of the 2d gauge field.
 Making field redefinition (see also \horts) $B= B' - f\del u, \  v= v' + \a'f
\tilde \theta$ we  can  transform  the model \modd\  into the
 flat space one.\foot{
The original  \sm  is  thus related to a flat space one  by a
combination of duality,  coordinate transformation and ``inverse" duality (cf.
\refs{\klt,\kk}).
If, however, the true starting point is the ``doubled" or ``gauged' model
\modd, then the transformation  to the  model corresponding to the flat space
is just a coordinate transformation on the extended configuration space of
$(u,v,r,\theta,\tilde \theta,B,\bar B)$.}
This transformation is not,  however, allowed  in  general
since it is not globally well defined. The dual coordinate
$\tilde \theta $ must have period $2\pi $ \rocver, so that $v'$
is  defined only if both $\phi=(v+u)/2$ and $t=(v-u)/2$
are assumed to be compact with period $2\pi R$, and if $f,R$ satisfy $\a'
f=2kR$, $k$=integer.

%%%%%%%%%%%%%%%%%%%%%%%%%%%%%%%%%%%%
\subsec{Point-particle limit  and  solution of tachyon equation}
%%%%%%%%%%%%%%%%%%%%%%%%%%%%%%%%%%%%%%%%%

The basic differential operator which dictates the propagation of
point-like (ground state) scalar particles
 in the  background  \lag\   and which should be
the zero-mode part of the Hamiltonian of the corresponding CFT is,
to leading order in the $\a '$-expansion, the
  Laplacian of the metric \metricone\ (the dilaton here is constant).
It has the simple form
\eqn\kg{ \Delta = - {1\ov{\sqrt{ -G}}} \del_\m ({\sqrt{ -G} }G^{\m\n} \del_\n)
= -4 \del_u\del_v  - [\del^i - 2 A^i(x)\del_v][\del_i - 2 A_i(x)\del_v] \ . }
The exact tachyon vertex operator is thus the solution of
 \eqn\tac{ \a' \Delta  T + O(\a'^2)T  = 4 T  \  .  }
The $O(\a'^2)$ corrections
  are scheme-dependent and may be non-vanishing in the scheme in
which
the exact expressions for the
\sm couplings (metric, dilaton and antisymmetric tensor)
do not depend on $\a'$ (for a discussion of $\a'$-corrections to the
tachyon equation see [56] and refs. there).
The exact form of the tachyon equation is usually hard to determine
without knowledge of the underlying conformal theory.
It turns out that in the present model one can actually fix its  form
 already at the \sm level
by using the relation (for $R\to \infty$) to the WZW model
of ref. [17].
Let us first omit the $\a'$-corrections in (2.33)
(we shall discuss how to incorporate them at the end of this
subsection).

 To relate the  operator \kg\ to the Hamiltonian of a point-particle limit of
our model let us
ignore the $\s$-dependence of the fields
in \lag\  ($v,u= \phi \pm t$)
\eqn\poi{
 I_{\rm particle}= {1\ov 4\a'} \int d\tau \big[
\dot u \dot v   + \dot x^i \dot x_i + 2A_i(x) \dot x^i \dot u
\big]\ . }
On the equations of motion $u=u_0 + p_v \tau, \ \dot u = p_v =\const, $
so that \poi\ becomes  very similar to the  standard  action   ( $  \int d\tau
\big[- \dot t^2    + \dot x^i \dot x_i + eA_i(x) \dot x^i
\big] $)
 of a charged particle  in an
external magnetic  field.
The peculiarity of the present
 model is that the charge $e$
is equal to  (twice)  the  ``light-cone" momentum $p_v$, or, equivalently, to
the sum of the  momentum
in the compact direction  (or ``Kaluza-Klein charge") $p_\phi$
 {\it and} the energy $p_t=E$,  $p_v= \ha ( p_\p + E) $.
The dependence on $E$  is  due   to the
 curvature of space-time (proportional to the magnetic field
energy density, cf. \eee).
The  Hamiltonian corresponding to \poi\
\eqn\hha{ H = 4\a' \{  p_u p_v +  \four [p_i  - 2  p_v A_i(x)]^2 \}  \ ,
 }
 directly leads  to \kg\ upon quantisation.
Representing the solution of \tac\
as (if $\p$ is periodic the integral over $p_u +p_v$ is actually a  sum)
\eqn\soo{
T(u,v,x) = \int dp_u dp_v \  \exp (i p_u u + ip_v v)  \   \td T (p_u,p_v,x) \ ,
}
we find that $\td T $ satisfies
\eqn\papp{
- [\del^i - ie A^i(x)][ \del_i -  ie A_i(x)]\td T  = \m \td T  \ ,  } $$   \  \
e\equiv  2p_v =p_\p + E \ , \ \ \ \ \ \
\m=  {4/\a' }  -  4 p_up_v  \  .  $$
This equation is formally the same as   the Schr\"odinger equation for a
charged particle in a magnetic field.\foot{One should keep in mind, of course,
that
the original equation we started with \kg\
is not a Klein-Gordon equation for a charged particle in flat space,  since it
contains an  extra $\del_t$-term  (in $\del_v$) on the usual place of the
charge. This is reflected in the presence of the energy in the   above relation
for the effective
charge $e$.  }

Eq. \papp\  can be easily  solved explicitly for
 the constant magnetic field \parr,
for example, in the  $D=4$ case of $F_{ij}= f \ep_{ij}$.
Using the gauge \parr, i.e. $A_i = - \ha f \ep_{ij} x^j$ one
 can put \papp\ in the form (we assume $fp_v > 0$)
\eqn\pappw{
2(C^\dagger C+ CC^\dagger) \td T  = \m \td T  \ , \ }
\eqn\ppw{  C^\dagger\equiv  -i(\del_x - \ha f p_v x^*)\ ,  \ \
C\equiv  -i(\del_x^* + \ha f p_v x)\  , \  \   \ x\equiv  x_1 + i x_2  \ , \ \
x^*\equiv  x_1 -i x_2  \ ,  }
\eqn\ann{ [C,C^\dagger] = f p_v\ , \ \ \  \
[b_0,b^\dagger_0] = 1\ , \  \ \ \ C \equiv  \sqrt{fp_v} b_0 \ . }
Thus the spectrum is given by
\eqn\spec{  4fp_v ( l + \ha ) = \m\ , \ \ \  \ \ \ \ l=0,1,2,...  \ . }
Defining also $  B^\dagger\equiv  -i(\del_x^* - \ha f p_v x)\ ,  \ \
B\equiv  -i(\del_x + \ha f p_v x^*), \ [B,B^\dagger] = f p_v$
(so that the angular momentum $J\sim B^\dagger B - C^\dagger C$)
one can represent the eigen-functions as:
$\psi_{l,n}= a_{l,n}(B^\dagger)^n(C^\dagger)^l \exp(-\ha fp_v |x|^2)$.\foot{
The final expression for $T$ may  have a simple integral representation
in terms of exponent of a bilinear form in $x$.
This is suggested by the observation that
in the case when $\p$ is non-compact we can use transformation
\coo\ to put the action into the plane-wave type form \fgf.
For the plane wave background \fgff\ the Klein-Gordon equation
is solved by \planetach\
$ T(u,v',y)=
  \int dp_vdp_i \hat  T (p_v,p_i)\exp\big[ i p_v v'  +  i{ (\a'p_v)\inv }u  + i
p_i y^i  -
\four  i p\inv_v   \int du g^{ij}(u) p_i p_j\big]. $
In our particular case of \fgf,
$ \int du g^{ij}(u) p_i p_j =  ( \sin u )\inv [ \cos u (p_1^2 + p_2^2) -2
p_1p_2 ] ,    $
and  $y^i$ are related to the original coordinates $x^i$
by the linear transformation \coo.  }

 Alternatively, we may  follow Landau \landau\ and  solve \papp\  in
 the gauge
$A_1=-fx_2,\ A_2=0$. According to \gau\ this  gauge is related to \parr\ used
above  by a shift of $v$ so that $v$ in the tachyon vertex \soo\ will be
related to $v$  in \moo\  by
 $v\to v  +  f x_1x_2$.
%\foot{Such shift of $v$ does not preserve
%the periodicity of $\p$ so the  models corresponding to the two gauges
% may  not  be   equivalent as string theories. This, however, does not matter
%for the problem of solution of the ``Schrodinger" equation \papp.}
After setting
\eqn\tat{
\tilde T=\int dp_1 e^{ip_1x_1} \Tau (p_u,p_v,p_1, x_2)
\ , }
eq. \papp\
 becomes the standard harmonic oscillator
differential equation,
\eqn\harmoni{
[-\del_2^2+\o^2(x_2-b)^2]\Tau =\m \Tau \ ,\ \ \ \ \o\equiv 2p_v f\ , \ \ \
 b\equiv -p_1/\o \ ,
}
with the solution
\eqn\sol{
\Tau (p_u,p_v,p_1, x_2)
=\sum_{l=0}^\infty \ c_l (p_u,p_v,p_1)\
  \exp [-\ha \o(x_2-b)^2] \ H_l[\sqrt{\o}(x_2-b)]
\ , }
where $\m$ (i.e. $p_u, \ p_v$, see \papp) and $l$ are
related by  the quantization condition equivalent to \spec\
\eqn\levels{
\m= 2\o (l+\ha) \ , \ \ \   \ \  {4/ \a'}- 4p_up_v =  4 p_v f (l+\ha) \ .
}
Setting   $ p_v= \ha (p_\p +E)$, $ p_u= \ha (p_\p - E)$,
where $p_\p = m/R, \ m=0,\pm 1,\pm 2, ...$ if $\p$ is
compactified on a circle\foot{Note that if $\p$ were not periodic we could
eliminate the dependence on $f$ by absorbing it into a rescaling of $E$ and
$p_\p$, $ \ p_\p +E\to f\inv (p_\p +E), \  p_\p -E\to f (p_\p - E)$.}
  we obtain the formula
for the tachyon energy\foot{Represented in terms of the dimensionless
parameters this equation becomes
$  -E'^2+ m^2  + (2l+ 1  )(m +E')h =4 r^2 \  , \  \ \ E'\equiv  R E\   , \ \  h
\equiv f R \ , \  \ r\equiv R/\sqrt {\a'}$.}
\eqn\enerver{
 -E ^2+ p_\p^2 +2(p_\p +E)f (l+\ha )=4/\a'\ .
}
As we shall see in Section 3.4 the expressions (2.41),(2.46)
for the tachyon spectrum  agree with the result
  which follows from the
corresponding conformal theory up to just one $\a'$-correction term.
It  is possible to understand the origin of this
correction term as  follows.
In general, the differential operator in \tac\ contains corrections of
the form $ (G^{\m\n} + c\a' H^{\m\l\r} H^\n_{ \ \l \r} + ...) D_\m D_\n
+ ... $. In the present model the non-vanishing component of $H_{\m\n\l}$ is
 $H_{uij} = - f \ep_{ij}$
and using its `null' structure it is possible to argue that
all other corrections
are absent
(assuming one uses the scheme based on dimensional regularisation).
Comparison with WZW \sm  shows that in the scheme
where the metric does not receive $\a'$-corrections $c= -1/4$ [56].
Alternatively, in the ``CFT scheme"
where the tachyon equation is not modified by $\a'$-corrections
the exact expression for the metric is given by
\eqn\exame{ G'_{\m\n} = G_{\m\n} - c\a' H_{\m\l\r} H_\n^{ \ \l \r}
= G_{\m\n} + \ha \a' f^2 \delta_{\m u} \delta_{\n u} \ , }
so that the exact \sm action
 receives  an extra term $\ha \a' f^2  \del u \bd u $
(and thus (2.34) also  gets  the correction $\ha \a' f^2 \dot u \dot u $).
The presence of such term is in agreement with the expression for the
stress tensor (or Hamiltonian) in the $E^c_2$
WZW theory \refs{\kk,\sfee}
 (because of the null structure of the $\a'$ correction term
in the stress tensor its presence is consistent with the ``free" value
of the central
charge). While in the  $E^c_2$
WZW model the $\a' \del u \bd u$ term in the action can be redefined away by
a shift of $v$ this is no longer possible  in the present model where
$u+v$ is compact.
 Irrespective of the interpretation (i.e. of the  scheme choice)
the tachyon equation thus contains an extra $2\a'f^2 \del_v^2$ term,
i.e. $\m$ in (2.37),(2.38) should be  shifted by $2\a'f^2 p_v^2$.
As a result, the $\a'$-corrected form of \enerver\ becomes equivalent
to the  CFT expression.

Combining \soo,\tat,\sol\ and
performing the integral over $E=p_v-p_u$ using \enerver\
we finish with
the  expression for the general solution  of \soo\
$T$ in terms of  the  integral over $p_1$ and the double sum over $l$ and $m$.
The tachyon vertex operator with given conserved quantum numbers
corresponds to a particular term
in this representation.

%\foot{The tachyonic vertex can be put in the form
%suitable for calculation of the scattering amplitudes
%Thus the  exact tachyon vertex operator can be written as
%\eqn\tachvet{
%T(u,v,x)=\sum_{l=0}^\infty \int dk_udk_1 dk_a c_l (e,k)
%e^{ik_uu+iev+ik_1x_1  +ik_ax_a} e^{-\o(x_2-b)^2}H_l[(x_2-b)\sqrt{2\o}]
%}
%Since $H_l$ is a polynomial and the exponent is quadratic in $x_2$, it seems
%%in %principle possible to
%derive explicit expressions for scattering amplitudes. This may be done by
%%%introducing a source term for $x_2$ and explicitly performing the gaussian
%%path %integral. We will not investigate scattering amplitudes in this work.

%%%%%%%%%%%%%%%%%%%%%%%%%%%%%%%%%%%%%%%%%%%%%%%%%%
\newsec{Light-cone gauge quantisation and spectrum }
%%%%%%%%%%%%%%%%%%%%%%%%%%%%%%%%%%%%%%%%%%%%%%%%%%

In this section we shall solve the classical string equations corresponding to
the $D=4$ theory  \moo\ and then quantise the model in the light-cone gauge.
This is possible to do explicitly for the  model \jkl\  in any $D$
due to  its special structure:
(i) for all the models \lag\ there are the infinite dimensional symmetries
$u'=u  + h(\t -\s), \
v'=v + g(\t + \s), $
implying the existence of the two conserved chiral currents
$ {\cal J}_+ = \del_+ u, \ {\cal J}_-= \del_- v  + 2 A_i(x) \del_- x^i$;
(ii)  the  Lagrangian \jkl\ or \moo\  is bilinear in $x^i$. The existence of
the light-cone gauge makes the unitary of the model explicit.

%%%%%%%%%%%%%%%%%%%%%%%%%%%
\subsec{Solution of the classical string equations}
%%%%%%%%%%%%%%%%%%%%%%%%%%%%%%%%%%%%%

Using Minkowski world sheet notation and introducing the  complex coordinates
$x=x^1+ix^2, \ x^*=x^1-ix^2$ one can represent the  $D=4$ Lagrangian  \moo\ in
the following form:
    \eqn\lagra{ L = \del_+ u \del_- v  + \del_+ x \del_- x^*
+ \ha if \del_+ u ( x \del_-  x^*  -  x^*\del_- x) +\del_+x^a \del_-x^a\ ,
}
where  we have added  $D_{tot} -4$  free scalar fields  $x^a$ ($a=3,...,D_{tot}
-1$)
which are needed to saturate the central charge condition.  It is easy to show
\horts\
that the model \lag,\maxw\ has the same central charge as the free one, so we
are  to choose $D_{tot} =26$.
The  free-field  contribution of   $x^a$ will be ignored  in the most of our
discussion.
The equations of motion are  then given by
\eqn\equa{\del_-\del_+ u=0 \ , \ \ \
\del_+ [\del_- v  +  \ha if ( x \del_-  x^*  -  x^*\del_- x)]=0  \ ,  }
 \eqn\equatwo{
 \del_+\del_-x^*=if\del_+u\del_-x^* \ ,\ \ \ \ \
\del_+\del_-x=-if\del_+u\del_-x \ .
}
Their  general solution is
\eqn\solus{
u= u_+ + u_-\ ,\ \  \    x= \exp (-if u_+) y\ , \
\ \ x^*= \exp (if u_+) y^*\  ,  \  \ \ y= y_+ + y_- \ ,}
\eqn\solutwo{
v= v_+ + v_-  + \ha  if ( y^*_+ y_- -  y^*_- y_+ )\ ,
}
\eqn\solutrhee {
\p = \ha (u+v) = \p_+ + \p_-   + \four if ( y^*_+ y_- -  y^*_- y_+ ) \ ,
}
where $u$ and $y$ satisfy the free wave  equations and
lower signs indicate dependence on $\s_+=\s + \t$ or $\s_- = \t-\s $, i.e.
$u_\pm = u_\pm (\s_\pm), \ v_\pm = v_\pm (\s_\pm),$ etc.
Thus $x, x^*$ are related to the free fields $y, y^*$
by an $SO(2)$ rotation.

\lr \dabh {A. Dabholkar, ``Strings on a cone and black hole entropy",
HUTP-94-A019,  hep-th/9408098; ``Quantum corrections to black hole entropy in
string theory",    hep-th/9409158;
D.A. Lowe and A. Strominger, ``Strings near a Rindler or black hole
horizon", hep-th/9410215.}

Let us now fix the  residual
  conformal symmetry $\s_\pm \to F(\s_\pm)$ by choosing the ``light-cone"
gauge:
\eqn\lcg{
 u = u_0+  p_+ \s_+ + p_-\s_- \ . }
Then  $x, x^*$ can be written as ($u_0$ is absorbed into the definition of $y$)
\eqn\xlcg{
   x= \exp (-if p_+\s_+) y, \  \ \ x^*= \exp (if p_+\s_+) y^* \ .   \
}
 The periodicity condition $x (\s + \pi,\tau) = x(\s,\tau)$ is solved if
$y$ satisfies  the ``twisted" boundary condition,
 $$y(\s + \pi, \tau)= \exp (ifp_+\pi) y(\s,\tau)\ , $$  implying
\eqn\yplumin{
 y_+ = \exp (ifp_+\s_+) z_+ \ ,  \ \ \ \ \  y_-= \exp (-ifp_+\s_-) z_- \ ,
}
where  $z_\pm$  can be expanded as follows,
\eqn\fourie{ z_+ =  i  \sqrt{\a'/ 2 } \sum_n \tilde a_n \exp (-2in \s_+)  \ ,
 \ \     z_- =  i  \sqrt{\a'/ 2 }
\sum_n a_n \exp (-2in \s_-) \ . }
The explicit form of the solution is thus
\eqn\xsol { x=  z_+ +  \exp (-2\tau ifp_+) z_- \ , \ \ \ \  x^*=  z_+ ^*+  \exp
(2\tau ifp_+) z_- ^*\ ,
}
\eqn\vzz {
 v= v_+ + v_-  +  \ha if [\exp (-2ifp_+\tau )   z^*_+ z_- -
\exp (2ifp_+\tau ) z^*_- z_+ ] \ ,}
\eqn\phisol {\p =\ha  (u+v)=  \p_+ + \p_-   +\four  if[\exp  (-2ifp_+\tau )
z^*_+ z_- - \exp (2ifp_+\tau ) z^*_- z_+ ] \ . }
 Since $\p$ is  assumed to be periodic with period $2\pi R$, its zero mode part
 will include the winding term.  The zero mode parts of the fields are then
\eqn\ze{  \p_{\rm zero} = \p_0   + q \s + s \t, \ \ \ \ t_{\rm zero}=t_0 + p\t
\ ,  \ \ \ q= 2Rw\ , \ }
\eqn\zer{  u_{\rm zero} =u = u_0 + p_+ \s_+ + p_-\s_- \ , \ \ \  p_\pm =
\ha(\pm q+ s -p) \ , }
\eqn\zee{\ v_{\rm zero} =  v_0 + q_+ \s_+ + q_- \s_-\ , \ \
\ q_\pm  = \ha (\pm q +  s +p)\ .  }
Here  $w$ is an integer winding number and  $s$ and  $p$ take continuous
values.
%\eqn\zee{  p_\pm = \ha(\pm q+ s -p)\ , \ \ \
 %q_\pm  = \ha (\pm q + s +p).  \ }

Above we assumed that momentum $p_+$ has generic value.
A special case is when $fp_+=2k, \ k=0,\pm 1 , \pm 2 , ... $.
In the sector with such special $p_+$
the field $y(\s,\t)$ satisfies the standard ``untwisted" boundary condition
and thus contains the translational part
$p_y \tau$ (absent for generic $fp_+$).
That means that for such  $fp_+$  (i.e. for special $p_+$ for a given
$f$ or for special magnetic field in each $p_+$-sector)
the ``$y$-string"  can move freely on the 2-plane.
The solution corresponding to this special case
can  still  be described   by \xsol\
if we rescale the oscillators $\tilde a_n= (n -\ha fp_+)\inv \tilde a'_n,
 \ \  a_n= (n +\ha fp_+)\inv a_n'$
and take the limit $fp_+ =2k + \ep, \ \ep \to 0$ (so that  $\tilde a'_k+
a'_{-k}
\sim p_y$).\foot{The solution  for $y$  is analogous
to string motion   on the $R^2/Z_k$ orbifold \dabh\
with generic solution representing  string modes of
the twisted sector (localised near the tip of the cone)
and the special
one -- the untwisted sector (strings that can move over  the plane). }

%%%%%%%%%%%%%%%%%%%%%%%%%%%%%%%%%%%%%%%%%%%%%%
\subsec{Stress tensor and Hamiltonian}
%%%%%%%%%%%%%%%%%%%%%%%%%%%%%%%%%%%%%%%%%%%%%
The non-vanishing
stress tensor components corresponding to the model \lag\  are  given by
%\eqn\tminus{
%T_{--}=\del_- u \del_- v  + 2A_i(x) \del_- x^i \del_- u +
%\del_- x^i \del_- x_i\  ,
%}
\eqn\tplus{
T_{\pm\pm}= \del_\pm u \del_\pm v  + 2A_i(x) \del_\pm x^i \del_\pm u + \del_\pm
x^i \del_\pm x_i\ . }
Computed on the classical solution
in the present case of \lagra,\solus\ they are
\eqn\tmin{
T_{--} = p_-\del_-v_-  + \del_- y_-\del_- y^*_-
+ \ha ifp_- ( y_- \del_- y^*_- - y^*_- \del_- y_- )  \ ,
}
\eqn\tplu{
T_{++} = p_+\del_+v_+  + \del_+ y_+\del_+ y^*_+
 -  \ha  ifp_+ ( y_+ \del_+ y^*_+ - y^*_+ \del_+ y_+) \ .}
The classical constraints $T_{--}=T_{++}=0$ are then easily solved  and, as
usual, determine
$v_\pm$  in terms of the free fields $y_\pm$ or $z_\pm$.
The classical expressions for the Virasoro operators $L_0, \tilde L_0$ are
\eqn\vira{
L_0={1\ov 4\pi\a' } \int_0^\pi d\s\  T_{--} =   {p_-q_- \ov 4\a' }
 +\ha\sum_{n }  \big(n+\ha {fp_+}\big)\big(n+\ha {fq}\big) a_n^{*} a_{n}\ ,
}
 \eqn\soro{
\tilde L_0={ 1\ov 4\pi\a' } \int_0^\pi d\s\  T_{++} =   {p_+q_+\ov 4\a'} + \ha
\sum_{n}  n\big( n-\ha  {fp_+} \big)       \td a_n^{*} \td a_{n} \ .
 }
Hence
the Hamiltonian $H=L_0+\td L_0$ is given by
% (cf. \hhh)
\eqn\hamilto { H= {  q^2 + s^2 -  p^2\ov 8\a' } +  \ha \sum_{n }  \big( n+ \ha
{fp_+}\big) \big( n+ \ha {fq} \big) a_n^{*} a_{n }+ \ha
\sum_{n}  n\big( n- \ha {fp_+} \big) \td a_n^{*} \td a_{n}
\ . }
The relation to the point-particle expression \hha\  is established by dropping
all $\s$-dependence (i.e. terms with $n\not=0$), setting winding number $q=0$
and expressing $s$ and $p$ in terms of conserved
momenta $p_u$ and $p_v$.

%%%%%%%%%%%%%%%%%%%%%%%%%%%%%%%%%%%%%%%%%%%%%%%%%
\subsec{Operator quantisation}
%%%%%%%%%%%%%%%%%%%%%%%%%%%%%%%%%%%%%%%%%%%
We can now quantise the  theory in  a
standard way by promoting the Fourier modes to operators acting in a Fock space
and demanding the canonical commutation relations,
\eqn\canon{
[P(\s ), x^*(\s')]=[P^*(\s ), x(\s')]= -i\delta(\s -\s')\   , \ \ [x^i(\s
'),\del_\s x^j(\s )]=0\ ,
}
where $P=\ha (P_1+iP_2), \  P^*=\ha(P_1-iP_2) $ are the momenta. Using that
$ P=  {1\ov 4\pi \alpha'} ( \del_\t x + i f p_+ x) $
%= {1\ov 2\sqrt{2\a'} \pi } \sum_{n} \bigg[
 %    (n- \ha {fp_+\ov  } )a_n^{+}  \exp (-2in \s_+)
%+ \exp(-2if p_+ \t)  (n+ {fp_+\ov 2} )a_n^{-}  \exp (-2in \s_-) \bigg]
%$$
we  find the relations
\eqn\fff{[ a_n, a_m^{*}] =    2  (n+ \ha {fp_+} )\inv   \delta _{nm}\ , \ \ \
[ \td a_n, \td a_m^{*}] =    2  (n - \ha {fp_+} )\inv   \delta _{nm } \ . }
 We can thus  introduce  the creation and annihilation operators:
\eqn\ope{ [ b_{n\pm}, b_{m\pm }^{\dag }] = \delta _{n m} \ ,\ \
 [\td b_{n\pm}, \td b_{m\pm }^{\dag }] = \delta _{n m} \ ,\ \  [b_0,b_0^{\dag
}]=1 \ , \ \ [\td b_0,{\td b}_0^{\dag }]=1 \ , }
\eqn\rightope{
 b_{n+}^{\dag }= a_{-n} \omega_- \ ,\ \  b_{n+}= a_{-n}^* \omega_-\ ,\ \
 b_{n-}^{\dag }= a_{n}^* \omega_+ \ ,\ \  b_{n-}= a_{n} \omega_+\ ,
}
\eqn\leftope{
 \td b_{n+}^{\dag }= \td a_{-n} \omega_+ \ ,\ \  \td b_{n+}=\td a_{-n}^*
\omega_+\ ,\ \
\td b_{n-}^{\dag }=\td a_{n}^* \omega_- \ ,\ \ \td b_{n-}=\td a_{n} \omega_-\
,\
}
 \eqn\ttt{
b_0^{\dag }=\ha \sqrt{fp_+} a_0^*  ,\  \ b_0=\ha \sqrt{fp_+}a_0
 , \  \ {\td b}_0^{\dag }=\ha \sqrt{fp_+} \td a_0  ,\ \ \td b_0=\ha
\sqrt{fp_+}\td a_0^*
\  , }
where $\omega_\pm \equiv \sqrt {  \ha \big( n \pm \ha {fp_+} \big) }$, \
$n=1,2,...\ $.
The subindices $\pm$ correspond  to components with spin ``up" and
``down" respectively. Above we have assumed $0<fp_+<2$. For $fp_+>2$
or $fp_+<0$ the  creation/annihilation roles of some operators are changed
  and the analysis
 remains essentially the same (see [57] for a detailed discussion of this
point).

In what follows we shall assume that all operators act on a particular
eigen-state of the zero-mode operators so that the latter
($q,s,p$, etc.) can be replaced by their eigen-values.
%We shall also adopt the  rule of symmetrisation (in $b^{\dag }, \ b$)
%of the classical expressions for the operators, with subsequent normal
%%ordering
%and defining possible infinite sums using $\zeta$-function regularisation
%To define the quantum operators we as usual symmetrize the classical
%expressions and then normal order.
The operators $L_0$ and $\td L_0$ will be normal ordered
with the
ordering constant being fixed by  the Virasoro algebra.
The free-theory normal ordering constant is shifted from 1 to
$1-\fourth \gamma (1-\ha\gamma) $ where $\gamma\equiv fp_+ $.
This corresponds to the use of the generalised $\zeta$-function
regularisation (see also [57])
 $$
  \sum_{n=1}^\infty (n + c )= \lim_{s\to
-1}\sum_{n=1}^\infty (n + c )^{-s}
= -{1\over 12}   + {1\ov 2}   c (1-c) \  .
$$
 Similar result is found in the
 open
string theory in  a constant magnetic field  \abo\ and is
typical to the case of   a free scalar field with twisted boundary conditions.
 We will see that this shift is also  consistent with the modular
invariant  path integral
approach discussed in Section 4.

Given that  $\p $ is a compact coordinate,  the zero mode part of its
canonical momentum  should be  quantized ($[p_\p, \p_0]=-i$),
\eqn\mom{
p_\p= \int_0^\pi d\s P_\p = { m\ov R}  \ ,\ \ \ \  m=0, \pm 1, \pm 2 , ... \ .
}
On the other hand,  using  eqs. \phisol, \xsol, \fourie, we  obtain
\eqn\pphih{
 p_\p={1\ov 2\pi\a'}\int_0^\pi d\s \big[ \del_\t\p +\ha if(x\del_-x^*-x^*\del_-
x) \big]=\ha {\a'}\inv  s+  f {\hat  J}_R\ .
}
Here
 ${\hat  J}_R$ is    the ``right" part of the angular momentum operator, which
after symmetrising the classical expression and normal ordering
has the form
\eqn\angulr{
{\hat  J}_R= - b^{\dag }_0 b_0 -\ha    +\sum_{n=1}^\infty \big(  b^{\dag
}_{n+}b_{n+} - b^{\dag }_{n-}  b_{n-} \big)= J_R - \ha
 \ . }
 Since our background is stationary, the string has  also conserved energy
operator
\eqn\epjr{
E=  \int_0^\pi d\s P_t=-{1\ov 2\pi\a'}\int_0^\pi d\s \big[ \del_\t t  +\ha
if(x\del_-x^*-x^*\del_- x) \big] = -\ha {\a'}\inv p- f{\hat J}_R\ . }
Inserting eqs. \rightope, \leftope , \pphih , \epjr \ in eqs. \vira , \soro\
and \hamilto\ we finally obtain   (we  include  also the contribution of the
extra
$D_{tot}-4=22$  degrees of freedom $x^a$)
   \eqn\lzero{
{\hat  L}_0=
\fourth  \big[\a'(-\E^2+\pp^2_a ) + ({m\ov r}-wr)^2\big]
- {1\ov 2} \a'  (Q + \E)\f { J}_R  + N -c_0 =L_0 - c_0         \ ,
}
\eqn\lzerob{
{\hat  {\td L}}_0= {1\ov 4}\big[ \a' (-\E^2+\pp^2_a) + ({m\ov r}+ wr)^2\big]
- {1\ov 2}\a' (Q+ \E)\f { J}_R  + {\td N}  -c_0 = {\td L}_0-c_0  \ ,
 }
\eqn\hamiltoni{
\hat  H =  {1\ov 2}  \big[\a'( -\E^2 +  \pp_a^2)  + w^2r^2 +{m^2\ov r^2} \big]-
\a' (Q+  \E)\f { J}_R
+ N+ {\td N}      - 2c_0 = H -2  c_0 \ , }
with the Virasoro conditions  $L_0=\td L_0=c_0$,
where
\eqn\defi{  r\equiv {R\ov \sqrt{\a' }}\ ,\ \
\ \ Q\equiv  {1\ov \sqrt {\a'}}  ({m\ov r} +  wr)= p_\p + {wR\ov \a'}  \ , \ \
c_0 \equiv 1- \four \gamma(1-\ha \gamma) ,
}
 $$
\gamma\equiv fp_+=\a' (Q+\E )\f\ ,
$$
and $N$ and $\td N$ are the   operators
\eqn\nnn{
 N= \sum_{n=1}^\infty n ( b^{\dag }_{n+}b_{n+}+ b^{\dag }_{n-}b_{n-}
+ a^{\dag }_{na} a_{na} ) \ ,\ }
\eqn\nn{
{\td N}= \sum_{n=1}^\infty n ( \td b^{\dag }_{n+}\td b_{n+}+ \td b^{\dag
}_{n-}\td b_{n-}+\td a^{\dag }_{na} \td a_{na}   )  \ . }
%To define $\hat N, \ \hat {\td N}$  we, as usual, symmetrized
% the classical expressions  and
%used the $\zeta$-function regularization ($\sum_{n=1}^\infty n= \zeta
%(-1)=-1/12$) assuming that $D_{tot}=26$.
Note   that ${\hat L}_0, {\hat {\td L}}_0, \hat H$
 depend on non-trivially on $f$ because of the periodicity condition in $\p$:
if $\p$  were  noncompact,   $f$ could  be eliminated
from \lagra\ by a coordinate transformation, or, equivalently,
could  be absorbed into continuous momenta and energy ($Q+E\to f\inv (Q+E), \
Q-E\to f(Q-E)$).
The Hamiltonian \hamiltoni\
of our CFT contains one $\a'$ correction term ($\g^2$ in $c_0$).
For $R=\infty$ it is in agreement with the Hamiltonian of the $E^c_2$
WZW theory which also contains one $1/k$ correction \kk.
This implies that the normal ordering (regularisation) prescription we
have used corresponds to the normal ordering used in the current
algebra  approach.

The above  operators and hence the  spectrum are  manifestly invariant under
the
duality transformation  $m\leftrightarrow w$,
$r\leftrightarrow 1/r$.
%It is interesting to note also  that
%(in agreement with \hhh) $H$ has only linear dependence on
 The field strength $f$     is coupled to
a combination of the energy with the ``left" momentum  $Q$ in the $\phi$
direction as in the  point-particle limit case  discussed in Section 2.5.
 The linear term $EfJ_R$ in $H$  coupling  the energy to the magnetic field
originates from
 the $A_i(x) \del_- x^i\del_+ t$  term  in the action \acti\ (which  is
necessary for conformal invariance and  represents the   deformation of the
geometry due to the magnetic stress-energy density).
Note also that only the right part of the angular momentum couples to the
magnetic
field (in particular, the zero mode operators corresponding to the left sector
do not appear in the Hamiltonian).

The correspondence with the particle  Hamiltonian  \hha,\pappw\
is established by  dropping all   oscillator terms, i.e.  keeping only the
zero mode part of ${\hat J}_R$ in \angulr\  (and setting
$w=0$),
\eqn\zerw{  \hat H_0 = {1\ov 2}\a'   \big[-\E^2+ { p}_\p^2  +  \pp_a^2
  +2 ( { p}_\p+  \E)\f (b^{\dag }_0 b_0 + \ha)
-\ha \a' f^2(p_\phi +E)^2 \big] -2       \ ,
}
and observing that
$b^{\dag }_0, \  b_0 $ directly correspond to the  creation/annihilation
operators in  \ann\ (note that  $p_v= \ha (p_\p + E)$, i.e. $p_v$
corresponds to $\ha p_+$,  cf. also \xsol,\ttt).
 The zero mode part
of ${\hat J}_R$ produces  the  Landau-type  term in   $H$ \hamiltoni\ (cf.
\spec)
\eqn\land{
H_{Landau}=  \a' (Q+\E)\f (b^{\dag }_0 b_0 + \ha) \  \to \
\a'  (Q+\E)\f(l+\ha )\ ,\ \ \ \  l=0,1,2,... \ , }
where $l$ represents  the
``right" part of the total orbital angular momentum.

 The    final Virasoro conditions are
\eqn\rotat{ L_0=\td L_0\ \  \to \ \
N-\td N=mw\ ,  }
\eqn\energia{ H=2c_0  \ \  \to \ \
  \k \big[ \E + {\k\inv \f }({\hat J}_R+\ha \a' Qf)\big]^2} $$  = -4{\a'}\inv
   + 4 \td N{\a'}\inv   +
{\k\inv } (Q - \f {\hat J}_R)^2 +\pp_a^2 \ ,
$$
  $$  \k \equiv 1+\ha \a'f^2
 \ .    $$
Here  ${\hat J}_R$ is the sum of the
``right" contributions to  the orbital  and spin momentum,
\eqn\jjj{ {\hat J}_R = -  b^{\dag }_{0}b_{0} -\ha  + S_R \to  -(l+ \ha)  + S_R
\ , \ \ \ \  S_R =
\sum_{n=1}^\infty \big(  b^{\dag }_{n+}b_{n+} - b^{\dag }_{n-}  b_{n-} \big)
 \ . }

Eq. \energia \  implies the following expression for the magnetic dipole moment
of an arbitrary
physical state: $\ \mu  =\sqrt {\a'}(1 + {Q/ M})  S_R ,   \ $
where $M$ is the rest mass of a  state. Hence the tree-level gyromagnetic
factor $g$ (defined in the frame corresponding to the stationary metric \did)
%(by definition, $<\mu>= g\sqrt {\a'}Q<J>/2M$, $J$ is the total angular
%%%momentum)
 is
given by
\eqn\magmom { g= 2(1+{M\ov Q} ){\langle S_R\rangle \ov \langle S\rangle} \ , }
 where $S$
is the total spin  (i.e. the sum of the left and right contributions).
  A similar  formula (without the $M/Q$ term)  was  obtained in ref. \rs \ for
the  model \actio\  which does not contain  the  coupling
$A_i(x) \del_- x^i\del_+ t$ (the expression of \rs\
turns out to be more universal than \magmom, for a detailed discussion
see  [57]).
  We thus confirm
the result of  \rs\ that    there
may be physical
states  with $g\neq 2$. This is a novel feature of  the {\it  closed }
string theory
 (all standard elementary particles  have $g=2$ as the  tree-level value
of the  gyromagnetic factor;
$g=2$ is  also true for the states of the
open string theory
in a magnetic field  \gtwo).
 See also a discussion in Section 6.

%%%%%%%%%%%%%%%%%%%%%%%%%%%%%%%%%%%
\subsec{Spectrum}
%%%%%%%%%%%%%%%%%%%%%%%%%%%%%%%%%%%
Physical states are constructed   by applying the creation operators (i.e.
$b^{\dag }_{n\pm}, \ \td b^{\dag }_{n\pm}$ and $ a^{\dag}_{n a}, \ \td
a^{\dag}_{n a}$)
to the Fock space vacuum $|0;\{m,w,l,p_a,E\} \rangle $, where $m$ and $w$
satisfy \rotat.
Henceforth we choose the frame in which $p_a=0$.
For  $f=0$ the spectrum  is the standard one  of  the bosonic closed string
theory with one dimension compactified on $S^1$. For a non-zero field $f$
the energy levels split according to the value of
the ``right" contribution to the angular momentum  $J_R$ and the
``left" momentum in the compact direction   $Q$.

The energy of the  ground state $|0;m$=$w$=$l$=$0\rangle $ is shifted  with
respect to the zero-field case
$\k  E^2 -  Ef =-4/\a'  ,\  \  \k = 1+\ha \a'f^2$.
It is always tachyonic.
For the
  general  scalar non-winding ($w=0$) ``tachyonic"  state   with  nonzero
$m$ and $l$ one   has
\eqn\enertach{
-E ^2+ Q^2 +2(Q +E)f (l+\ha )-\ha \a' f^2(Q+E)^2=4/\a'\ , \ \ \ \  \   Q=
p_\p=m/R \ .
}
This is the same
expression as  \enerver\ we  have found by directly solving the
tachyon equation \tac\  in  Section 2.5.  Since $Q$ can take both positive and
negative values there are tachyons for any $f$.
At the same time, the energy levels of all the states with  $Q=0$
%\eqn\qsoscer{Q=0}
are also modified by the presence of the magnetic field.
  As was already mentioned,
the  modification of the spectrum  as compared to the zero-field case  can be
understood as being due
 to  the  presence of  the  curved metric and the antisymmetric tensor
backgrounds induced by the magnetic field.

Let us now discuss some particular states.
First, consider the zero-energy states with  $N=\td N=1$.
For example, a   component of the ``photon" with $S_R\neq 0$ is represented by
 $b^{\dag }_{1\pm}\td a^{\dag }_{1(D_{tot}-1)}|0;m$=$w$=$0
\rangle $.
The ``graviton" components  correspond to
$b^{\dag }_{1\pm}\td b^{\dag }_{1\pm}|0;m$=$w$=$0\rangle $ and
$b^{\dag }_{1\pm}\td a^{\dag }_
{1a}|0;m$=$w$=$0\rangle$.\foot{Note that  with  our choice of light-cone
gauge on $u=\p-t$
  we have gauged away the oscillators
of the  compact ``internal" coordinate
 $\p $  instead of  removing the oscillators of the
longitudinal component $x^{D_{tot} -1}$. As a result,  some of the
states are represented
in a   way which is different from  their usual light-cone form.}
For these photon and graviton components one has $J_R=\pm 1 -l$.
Since $N=\td N=1$ one learns from
eq. \energia\ that $E=0$ for any value of the Landau level $l$.

 Next, let us look at   the states which
complete the $SU(2)_{R}$ massless vector
multiplet in the $f=0$  theory
compactified at the self-dual radius $r=1$. The components with $S_R\neq 0$ are
given by
$
b^{\dag }_{1\pm }|0;m=w=1\rangle  ,\ \ b^{\dag }_{1\pm }|0;m=w=-1\rangle .$
For them $\td N=0$,  $\ J_R=\pm 1 - l $, $\ l=0,1,2,..., $
and the  energy
\eqn\sutwor{
  \k \big[ E + {\k\inv f }({\hat J}_R+\ha \a' Qf)\big]^2 = -4{\a'}\inv   +
{\k\inv } (Q - \f {\hat J}_R)^2  \ , }
\eqn\stwo{  \ \ \ Q= {1\ov \sqrt{\a'} }
(r+ r\inv ) \ .
}
At the self-dual radius, $r=1$, an infinitesimal magnetic field $f>0$ renders
the component with $J_R=1$   tachyonic. This instability is the same one as
in the
 non-abelian gauge  theory  \refs{\salam , \nielsen}. Away from the self-dual
radius, this state has positive energy for small $f$ and becomes tachyonic at
 some critical magnetic field.
The states corresponding to the $SU(2)_L$ multiplet   have positive $E^2$
 and $E^2=0$ at the self-dual radius (the  state $b^{\dag }_{1\pm
}|0;m$=$-w$=$-1\rangle $ for which
$\td N=1$ and $J_R=-l$  has $ E^2=0$ at $r=1$
regardless  of the value of $f$).

The effect that the magnetic field produces on the energy of a generic
state is a combination of the gyromagnetic interaction and the influence of the
space-time geometry.
The states which are more affected
are those with a  maximum value of $J_R$. These are the states on the
 first Regge trajectory, {\it viz.} states of the form
$(b^{\dag}_{1+} )^k \tilde P|0;l$=$0\rangle $, where $\td P$ is a
  product of left-moving creation operators. Such states have
$N=J_R=k$, $\ \td N=k - mw $. As one  gradually increases the magnetic field,
the energy decreases until a critical point at which it vanishes.
For  $k>>m,w$ this occurs when (see eq. \energia )
\eqn\zeroen{
{\a' }  Qf\cong {1\ov k-\ha}[\ha (wr-{m\ov r})^2+2(k-1) ] \ .
 }
Let us consider, for example, the case $r=1$ and $m=w$. Then we have
$\a'Qf\cong 2(k-1)/(k-\ha )$.
As the magnetic field is increased from ${2/(\a' Q)}-\epsilon $ to
${2/(\a'Q)}$ an infinite number of zero-energy excitations appear
progressively. Similar effect happens for generic  values of $r$.
On the other hand, instabilities will be expected when the energy will
develop an imaginary part, that is, when $\big[ E+{\k\inv f
}(\hat J_R+\a' fQ)\big]^2<0$. From eq.
\energia\  we see that this can only happen for
$\tilde N=0$, i.e., when there are
 no left-moving excitations, and  for $f$
greater
than the critical value,
\eqn\critic{
 f_{\rm cr}= {Q\hat J_R - \sqrt{{4\ov \a'}(\hat J_R^2-2)+2Q^2} \ov \hat
J_R^2-2}
\ ,\ \ \ \  mw=k>0\ .
}
 There will be infinitely many tachyonic instabilities
for an arbitrarily small   value of the magnetic field. Indeed, in the example
with $r=1$ and $m=w$ states, there are an infinite number of tachyons
with $\td N=0, N=m^2, \ J_R\cong 2m/(\sqrt {\a'} f) $.
The critical magnetic field for a given state   will be the lower,  the
larger  the value of its charge.

The fact that in closed string theory a finite magnetic field already produces
an infinite number of tachyonic instabilities confirms the picture suggested in
ref. \thermal \ on the basis  of the linear in $f$ approximation in the
simplified model
(2.1).  This   conclusion is different from what happens in the open string
case \refs{\abo,\ferrara}
 where  a finite magnetic field  can  lead  only to a finite
number of tachyons.
We shall further comment on this instability of a constant magnetic field
background in closed string theory    in Section 6.

%%%%%%%%%%%%%%%%%%%%%%%%%%%%%%%%%%%%%%%%%%%%%
\newsec{Partition function on the torus}
%%%%%%%%%%%%%%%%%%%%%%%%%%%%%%%%%%%%%%%%%%%%%%%

Below we shall demonstrate  that the results derived in previous section
using the
light-cone gauge operator approach  are consistent  with the one-loop (torus)
partition function $Z$ calculated in the path integral formalism.
We  will reproduce  the standard relation
(see e.g. \refs{\polch,\ginsp}) between the path integral and
operator approach expressions for $Z$ and  check its  modular invariance and
duality invariance in the compact target space direction.\foot{Both
symmetries must of course be present  in the partition function since
our model is based on  a $2d$ diffeomorphism invariant and duality invariant
world-sheet action.}
%%%%%%%%%%%%%%%%%%%%%%
\subsec{Path integral derivation}
%%%%%%%%%%%%%%%%%%%%%%%%%%%%%%%%%
We shall take the metric of the torus in the form
\eqn\me{ ds^2 = g_{\a\beta} d\s^\a d\s^\beta\ , \ \
 0<\s_\a\leq 1\ , \ \  \t=\t_1 + i \t_2 \  ,  }
$$
g_{\a\beta}=\left(\matrix {1&\t_1\cr \t_1& |\t|^2 \cr}\right)\ ,\ \  \
\sqrt g g^{\a\beta}= \t_2^{-1} \left(\matrix {|\t|^2&-\t_1\cr -\t_1& 1
\cr}\right)\  .
$$
All the  fields are periodic in $\s_\a$ with period 1.\foot{The correspondence
with the Euclidean  world sheet notation used in Section 2
is established  by setting $\t=i $ and interchanging  $\s_1$ with $\s_2$,
cf. footnote 2.}

The  Euclidean action  and the torus partition function for  our model \moo\
are given by
\eqn\ac{I={1\ov 4\pi \a'}  \int d^2\s \sqrt g [ \del_\a u \del^\a v +
\del_\a x^i \del^\a x_i
 + f (g^{\a\beta} +  i{\epsilon^{\a\beta}\ov \sqrt g } )
 \ep_{ij} x^i \del_\a x^j  \del_\beta u ] \ ,  }
\eqn\zee{  Z= \int [d^2\t] \td Z  (\t) \
, \ \ \  \ \ \td Z =  \int  [du][dv][dx^i]\   \exp(  -I  )\ . }
Integration over $v$ produces a $\delta$-function that constrains
$u$ to be a zero-mode of the Laplacian on the torus.
If $\p=\ha (u + v) $ is
periodic  with period $2\pi R$  it should
satisfy the  conditions
$\p (\s_1 +1, \s_2 ) = \p (\s_1, \s_2) + 2\pi R w, \
\p (\s_1 , \s_2 + 1 ) = \p (\s_1, \s_2) + 2\pi R w' , $ where $w,w'$ are two
integer  winding numbers. Then
 the zero modes  of $u$ and $v$ on the torus are given by
\eqn\ze{ u_*= u_0 + 2\pi R(w \s_1 + w' \s_2)\ , \ \ \
 v_*= v_0 + 2\pi R(w \s_1 + w' \s_2)\ . \  }
 After the integration over $v$ and $u$   the action in \zee\ is replaced by
 $I=I_* + I'$,  where
\eqn\act{    I_*= {1\ov 4\pi \a'}  \int d^2\s \sqrt g  \del_a u_* \del^a v_*  =
\pi  r^2 \t_2\inv { (w'-\t w)(w'-\bt w)}  \ ,  \ \ r\equiv R/\sqrt{\a'} \ , }
\eqn\acc{I'=   {1\ov 4\pi \a'}  \int d^2\s  \t_2^{-1} \big[(\del_2 -\t
\del_1)x^i
(\del_2 -\bt \del_1)x_i + f \ep_{ij} x^i (\del_2 -\bt \del_1)x^j (\del_2 -\t
\del_1)u_*\big]\ . }
The integration over the two coordinates
 $x^i$  gives  the following contribution to
$\td Z$  in \zee\
\eqn\de{ \td Z_x=  [{\rm  det'}( \Delta_0 + {\cal D} ) {\rm  det'}(  \Delta_0 -
{\cal D} )  ]^{-1/2} , \ \ \ \Delta_0\equiv  - (\del_2 -\bt \del_1) (\del_2 -\t
\del_1)
\ ,  }
\eqn\qu{   {\cal D}\equiv i f(\del_2 -\t \del_1)u_*
 (\del_2 -\bt \del_1) =  2\pi i h  (w'-\t w) (\del_2 -\bt \del_1) \ , \ \ \
h\equiv fR =\sqrt{\a'} fr \ .  }
Expanding $x^i$ as
\eqn\exps{ x^i= x_0^i + \sum_{(n,n')\not=(0,0)} a^i_{nn'} \exp 2\pi i(n\s_1 +
n'\s_2)\ ,  }
we get
\eqn\dee{ {\rm  det'}( \Delta_0 + {\cal D} ) =  {\rm  det'}( \Delta_0 - {\cal
D} )= \prod_{(n,n')\not=(0,0)}
{\pi^2\ov \t_2^2} (n' - \bt n)[ n'- \t n  +  h (w'-\t w)]  \ .  }

\def\tw{\tilde w}

As a result, we can represent \de\ as the standard contribution of the two
scalars times the  ``correction" factor $Y^{-1}$
\eqn\dde{ {\td Z}_x = [{\rm  det'}\Delta_0 ]^{-1}  Y^{-1}  \ ,  \ \ }
\eqn\ddd{
{\rm  det'}\Delta_0 = \t^2_2 \eta^2 \bar \eta^2 \ ,
\ \   \eta(q) \equiv  q^{1/24} \prod^\infty_{n=1} (1-q^n) \equiv  q^{1/24} {
f_0} (q) \ , \ \ q=e^{ 2\pi i \t} \ , }
\eqn\yy{
 Y  \equiv    \prod_{(n,n')\not=(0,0)}
[ 1 +  h {w'-\t w\ov n'- \t n}]   \ .   }
The non-trivial part of the  $\t$-integrand  in  \zee\
which contains the dependence on $R$ and $f$  is given by
 the sum over windings  of the product of the
semiclassical factor (exponential of the  classical zero mode a
ction $I_*$ in
\act\ and  $Y^{-1} $ (we include also the  standard zero-mode f
actor $r$)
\eqn\non{ W(r,h;  \t, \bt)= { r}
\sum_{w,w'=-\infty}^{\infty}
 \exp [- \pi r^2  \t_2\inv { (w'-\t w)(w'-\bt w) }] } $$\times
\prod_{(n,n')\not=(0,0)}
[ 1 +  h {w'-\t w\ov n'- \t n}]^{-1}\ . $$
 This representation makes it explicit that the
partition function  is  modular invariant for arbitrary $r$ and $h$
(as in Section 3 we add  22 extra free scalar  degrees of freedom to satisfy
the zero central charge condition),
\eqn\full{
Z = c_1\int d^2\t \   \t_2^{-14}
 e^{4\pi \t_2} |{ f_0}(e^{2\pi i \t})|^{-48}  W(r,h;  \t, \bt)\
 . }
In fact,  the transformations $\t\to \t +1 $ and $\t\to -1/\t$
are  ``undone"
by
the redefinitions of the summation parameters: $w\to -w', \ w'\
\to w,
\ n\to -n', \ n'\to n$.

\def\tw{\tilde w}

To  find the explicit representation for  $Y$
 one should be careful to use a
reparametrisation invariant (modular invariant)
regularisation of the
infinite product in \dee.
Let us first note that \eqn\yyy{
U(\t,\bar \t, \chi)\equiv    \prod_{(n,n')\not=(0,0)}
(n'-\bar \t n) (n'- \t n + \chi )
} $$  = \prod_{k\not=0}
k( k  + \chi )  \prod_{ n\not=0, n'}
 (n'-\bar \t n) (n'- \t n + \chi )\ . $$
Summing first over $k$ and $n'$ using that
$$\prod_{n=-\infty}^\infty (n + \chi) = \chi  \prod_{n=1
}^\infty (-n^2) ( 1 - {\chi^2\ov n^2}) =
2i \sin \pi \chi\ , $$  defining the product
$\prod_{n\not=0} \exp (i\pi \bar \t n ) \exp (-i\pi \t
n + i \pi \chi
)$ as $ \exp [ 2\pi \t_2  \sum_{n\not=0} (n + i \chi/2 \t_2)]
, $
and using the generalised $\zeta$-function regularisation
as in sect. 3.3 we finally obtain
\eqn\yyy{ Y\equiv {U(\t,\bar \t, \chi)\ov U(\t,\bar \t, 0)}
=
\  e^{{\pi  \chi^2
\ov 2 \t_2}} \
  { \sin \pi  \chi
 \ov  \pi  \chi}
\prod_{ n=1}^\infty {  (1 - \r\inv
   q^{n })(1 -  \r
   q^{n }) \ov (1 -  q^{n })^{2}}  }  $$
 = \   e^{{\pi  \chi^2
\ov 2 \t_2}}
 \  {\theta_1(\chi| \t)
\ov  \chi \theta'_1 (0| \t) }  \  , $$
where
\eqn\defi{ \tw \equiv w'-\t w \ , \ \ \ \  \r\equiv \exp ({2\pi i
h\tw}) \ , \ \ \ q=\exp({ 2\pi i  \t})\ ,  \ \chi\equiv h \tw\ .
 }
The resulting expression for $Z$ \full\ is manifestly
modular invariant
and will be shown below to be in agreement with the
operator formalism.

%%%%%%%%%%%%%%%%%%%%%%%%%%%%%%%%%%%%%%%%%%
\subsec{Equivalence with  operator formalism result }
%%%%%%%%%%%%%%%%%%%%%%%%%%%%%%%%%%%%
In  the operator formalism,  the one-loop  partition function
is obtained by using the Hamiltonian to propagate the  states  along  the
cylinder and
identifying its ends
\eqn\zoper{
Z=\int {d^2\t\over {\rm \t_2 } }\int dE\prod_{a=1}^{22} dp_a
\sum_{m,w=-\infty}^\infty  \Tr ( q^{\hat L_0} {\bar q}^{\hat{\td L}_0} )\ ,
}
where $q=\exp(2\pi i \t)$ and $ \hat L_0$ and $\hat{\td L}_0$ are the Virasoro
operators
constructed
in Section 3  (see \lzero,\lzerob)
\eqn\viras{
\hat  L_0=\fourth\a '[-(E+ f{\hat J}_R)^2 + p_a^2]-1-{1\ov 8}\gamma^2+ {\cal
L}_0 \ ,  \ \
 }
$$ \hat{ \td L}_0=\fourth\a '[-(E+ f{\hat J}_R)^2 + p_a^2]-1-{1\ov
8}\gamma^2+\td {\cal L}_0 \  ,
$$
\eqn\callo{
 { {\cal L}_0} =
\fourth  (mr\inv +   w r - \sqrt{ \a' }f{\hat J}_R)^2 -  mw +  N \ , \ \
\td {\cal L}_0 = \fourth (mr\inv +   w r- \sqrt{\a'} f{\hat J}_R)^2+\td N\ .}
  %\eqn\uuu{
%\sqrt {\a'} Q \equiv mr\inv +   w r  \ . }
It is convenient to express the exponential factor containing $\gamma^2 $
in the following way:
$$
\exp{(\ha  \pi \t_2 \gamma^2})=\sqrt{\t_2}\int dx \  \exp ( -\ha \pi \t_2
x^2-\pi
\t_2 \gamma x) \ .
$$
The  term  $\pi \t_2\gamma x$    can be absorbed into
a redefinition of $\hat J_R\to \hat J_R'\equiv \hat J_R-\ha x $, as can be
easily verified.
Integrating over $E, p_a$ we get \zoper\ with an extra  measure factor
$\t_2^{-23/2}$ and the same trace over the oscillator states
with $\hat L_0, \hat {\td L}_0$ replaced by  ${\cal L}_0, \td {\cal L}_0$.
 The next step is
to perform  the  Poisson resummation
(or  duality transformation on the world sheet), trading the sum over the
discrete loop momentum $m$
for the sum over the conjugate winding number    $w'$:
$$  \sum_{m =-\infty}^{\infty} F(m) = \sum_{w'=-\infty}^{\infty} \int ds\  F(s)
\ \exp(2\pi i s w') \  ,  $$
\eqn\poiw{
 F(m) = \exp [  \ha \pi i (\t-\bar \t)
 ( mr\inv + wr  -  \sqrt {\a'} f\hat J_R')^2   - 2\pi i  mw  \t ]  \ . }
Integrating over $s$ we get (up to  a numerical factor)
\eqn\fini{
 {r\ov \sqrt{\t_2}} \exp [-{\pi r^2 \ov \t_2} { (w'-\t w)(w'-\bt w) }]  \
\Tr \{ q^{ N } \bar q^{ \td N} \exp [2\pi i
 (w'-\t w)  r\sqrt {\a'} f \hat J_R']
\} . }
The first exponential factor is just the  standard semiclassical   contribution
 in \non. The
trace produces the  non-trivial magnetic field dependent contribution (see
\nnn,\jjj)
\eqn\traza{
 X ( q, \bar q, \tw h)\equiv \sum_{l=0}^\infty \Tr  \{ q^{ N} {\bar q}^{\td N}
\exp [2\pi i  \tw h  ( -  l  - \ha -\ha x + S_R) ] \}} $$ =  X ( q, \bar q, 0)
 e^{-\pi i h\td w x} {\pi h\tw \ov \sin (\pi h\tilde\o ) }
 \prod_{n=1}^\infty {(1- q^n)^2
\ov (1-\rho  q^n)(1-\rho ^{-1} q^n)}\  ,
 $$
which, after integrating over $x$,  is equal to $Y\inv$ (see \yyy ).
The non-trivial zero-mode factor of $h\tw$ (necessary for the
regular  $f\to 0$ limit)
originates from the correct normalisation
of the trace consistent with the zero-field limit: note that in \ann\ or \ttt\
one made a rescaling of the original $x$-field modes by the factor of $fp_v$ or
$\ha fp_+$ or, in the present setting,
$f (\del_2 -\t \del_1)u_*=f \tw$ (see \ze).
The zero mode normalisation condition  also makes the trace in \traza\
implicitly
$\t_2$-dependent, leading  to
an extra $1/\t_2$ factor in $X(q,\bar q, 0)$ (see e.g. \nahm\  for a discussion
of
 related  normalisation   issues).

We conclude that the path integral and operator approaches give indeed
equivalent
expressions for the one-loop partition function depending on
radius $R=\sqrt {\a'} r$ of
the  compact dimension and
the magnetic field $f$  (see \full,\yy )
 \eqn\zfinal{
Z(R,f) = c_1 \int d^2\t \   \t_2^{-14}
 e^{4\pi \t_2} |{ f_0}(e^{2\pi i \t})|^{-48} \  W(R,f; \t, \bt)\ , }
\eqn\ppp{ W=    {R\ov \sqrt {\a'}}
  \sum_{w,w'=-\infty}^{\infty}
 \exp\big[- {\pi R^2 \ov \a' \t_2}
{ (w'-\t w)[(w'-\bt w) +\ha \a' f^2(w'-\t w)]}\big] }
$$
\times
 {fR (w'- \t
w)\theta'_1 (0| \t)\ov \theta_1[  f R (w'- \t w)| \t] } .
 $$
The path integral derivation makes explicit the modular invariance
of $Z$ (see \non), while
 the operator derivation  demonstrates its duality invariance ( $Z$ in \zoper\
is manifestly  invariant
under $r\to 1/r$, $m\to w$).
In the limit $f=0$  \zfinal\ reduces to the standard
partition function of the bosonic string with one compact dimension.

In the limit of the non-compact $\p$-dimension, $R\to \infty$,
the dependence on $f$ disappears  and the partition function becomes equivalent
to the flat space one.\foot{This agrees with the observation made in \kkl\
 that
  the torus partion function for the model
of \napwi\ is equal to its flat space  value.}
In fact, the duality invariance of $Z$
implies that $R\to \infty$ limit is equivalent
to $R\to 0$. In  the latter case $h\to 0$ so that  $Y\to 1$ (see \yy)
and the final expression for $Z$ is the usual flat space one, i.e. a constant
times $R\inv\to \infty$ (diverging as a volume).
%\foot{
%A closely  related limit is that of an infinite magnetic field (for fixed $R$)
%$f\to \infty$: here again $h\to \infty$ but the exponential factor in $W$
%%%survives.
%Since there is no $f\to 1/f$ duality it is somewhat surprising to find that
%%the
%limit $f\to \infty$ gives the same free theory result.}

The partition function (4.25) (see also (4.26) below)  has
``extra" poles
at rational values of $h=fR$.
These singularities
 originate (upon integrating over $E$) from contributions of the extra zero
modes
(associated with a translational invariance in the $x,x^*$-plane, see Section
3.1)
that appear in  the string solution (cf. (3.11))
in a sector with
$p_+ f=2n$, i.e. $\a' f(E+Q)=2n$ ($n=$integer).
%The corresponding divergence can be projected out or
% regularised, e.g.,
%by defining $Z(R,f)$ by an analytic continuation from the region
%of generic $h$.
 A possible physical role of these
special $fR$ points  (or special values of momentum $p_+$ of a free  string)
deserves further clarification.

%%%%%%%%%%%%%%%%%%%%%%%%%%%%%%%%%%%%%%%%%%
\subsec  { Tachyonic instabilities}
%%%%%%%%%%%%%%%%%%%%%%%%%%%%%%%%%%%%

The  magnetic instability of our background  (implied by the presence of
tachyons  in the spectrum
found in Section 3)
is reflected  in extra singularities (or imaginary parts) in  $Z$.
This can be seen explicitly  from  the behavior of the integrand
for large $\t_2 $ if one starts
 from the operator formalism  representation \zoper\ and performs the trace
before integrating over $E$. One finds
\eqn\ztrac{
Z=\int  {d^2\t\over \t_2 ^{13} }\int dE \sum_{m,w=-\infty}^\infty
e^{\pi\t_2[4+\a'(E^2-Q^2)+\ha \gamma^2]-2\pi i\t
mw} |{ f_0}(e^{2\pi i \t})|^{-48} }
$$
 \times\  {\l \over {\rm sinh}\ \l } \ \prod _{n=1}^\infty (1- q
^n)^{2}[(1-e^{2\l } q^n)(1-e^{-2\l }q^n)]^{-1}
$$
where $\l \equiv\pi \t_2 \a '(Q+E)f=\pi \t_2 \g, \ \  Q\equiv  ( m/r +
wr)/\sqrt{\a'} $.
Expanding the different factors in powers of $q$ and $\bar q$, one gets
a sum in which each term represents the contribution of a corresponding  state
in  the spectrum.
% As can be seen directly from
%eq. \zoper, where the dependence on $\t_1 $ is given by
%$e^{2\pi i\t_1 (N-\tilde N-mw)}$, the role of the integral over $\t _1 $ is to
%%%impose the Virasoro constraint eq. \rotat\ .

In Section  3 we have seen that the energy of a state   is more affected
by the magnetic field (relative to the $f=0$ energy) the higher is
the  corresponding  value of $J_R$. In eq. \ztrac \ the states with highest
angular  momentum
for  a given level     correspond to the
terms obtained from the  leading terms in all the factors
except  $ (1-e^{-2\l } q)^{-1}$. They correspond to  the $\tilde N=0$
states discussed in  Section  3, i.e. states on the first Regge trajectory.
We can represent their contribution to  $Z$ as follows
\eqn\expanss{
Z=  \sum_{m,w=-\infty}^\infty \int dE
\int d\t_1 e^{2\pi i\t_1(k-mw)}
\int  {d\t_2\over \t_2 ^{13} } \sum_{k=0}^\infty e^{-\pi\t_2 M^2_k}+...\ ,
}
with
$$
M^2_k=-4-\a'E^2+{m^2\over r^2}+w^2r^2-2\a'(Q+E)f(k-\ha )+2k -\ha \gamma^2\ .
$$
Integrating  over $\t _1$ and  $E$ we  obtain
\eqn\ultim{
Z\sim  \k^{-1/2}\sum_{m,w=-\infty}^\infty
 \int  {d\t_2\over \t_2 ^{27/2} } \sum_{k=0}^\infty e^{-\pi\t_2
{M_k'}^2}\delta_{k,mw}+...\ ,
}
$$
{M_k'}^2=-4+{\k\inv}[{m\ov r}+wr-\sqrt{\a '}f(k-\ha )]^2\ , \ \ \ \k=1 +
\ha\a'f^2 \ .
$$
The partition function has  new divergences when ${M_k'}^2<0$. As expected,
this
occurs
exactly at the critical magnetic field \critic\  when  the energy develops an
imaginary
part.

 %%%%%%%%%%%%%%%%%%%%%%%%%%%%%%%%%%%%%%%%%%%
 %%%%%%%%%%%%%%%%%%%%%%%%%%%%%
\newsec{Heterotic string generalisations}
%%%%%%%%%%%%%%%%%%%%%%%%%%%

\def \hl {{ \l}}
%%%%%%%%%%%%%%%%%%%%%%%%%%%%%%%%%%%%%%%%%%%%%%%%%%%
Let us now  follow  Section 2.1
and address the question of how to embed  a constant
magnetic  field background in  a  closed superstring or heterotic string
theory.
Since there are no fundamental gauge fields in the type II
superstring theory  the only option is to couple
the gauge field to a compact   Kaluza-Klein coordinate
as in the bosonic case. The corresponding world sheet theory is then
the  direct  $(1,1)$ supersymmetric
extension of the bosonic \sm \lag.

As for the heterotic string theory, here there are more options.
One may  consider the $(1,0)$  and  $(0,1)$
 supersymmetric extensions of the
bosonic model (and to add  interactions in the  internal fermionic sector if
necessary for conformal invariance)   thus  getting the
magnetic  field again from the Kaluza-Klein sector (Section 5.1).
Another possibility
is to embed  the magnetic field directly into
the heterotic  gauge group, i.e. to couple $A_i$ to internal fermions.
This second approach  is more appealing  from a ``phenomenological"
viewpoint since in this case the corresponding $\s $-model background
will describe the  effect on the geometry when the usual
(i.e. coming from   $U(1)_{\rm e.m.}$)  magnetic field
 is turned on (cf. ref. \rs ).

If one  starts with   the   flat-space heterotic string
action {\gro}\
 and introduces the magnetic field
in   the internal fermionic sector
 \eqn\het{I=  {1\ov \pi \a'}
\int d^2 \s \big[ - \del t \bd t + \del x_i \bd x^i
 -   \l_{L}^t \del \l^t_L +    \l_{Li}\del \l^i_L }
$$  + \
 \psi_{RI} (\delta^{IJ} \bd + V^{IJ }_i \bd x^i) \psi_{RJ}
- \ha  {\cal F}^{IJ}_{ij}(V)  \psi_{RI} \psi_{RJ} \l_L^i\l_L^j \big] \  , $$
where  $V^{IJ}_i= T^{IJ} A_i$
($T$ is proportional to a $U(1)$ generator
of  the  gauge group),  then, as in the bosonic case  \actio,
such model will not be conformally invariant
since the magnetic field will deform the space-time geometry.

To find  a  conformally invariant  modification  of \het\
one may  consider  $\psi_R$ as playing the role of the internal Kaluza-Klein
field $\p$
of the bosonic model \la,\lagr.\foot{A different conformal embedding of
a monopole-type magnetic field into the gauge sector of the  heterotic string
theory
which uses  $SU(2)$ WZW model  was recently discussed in
\kkkk.}
Assuming that the $\p$-terms of the second line in \lagr\  are ``fermionised"
and supersymmetrising (in the  $(0,1)$ way \refs{\sen,\hulw})
 the $t,x^i$-terms of the first line in \lagr\  we
obtain  the following heterotic \sm  action
\eqn\hett{I' = {  {1\ov \pi \a'} }
\int d^2 \s \big[  -  (\del t  + A_i  \del x^i)(\bd t  +  A_i  \bd x^i)
 +   \del x^i \bd x_i +  A_i  (\del x^i \bd t - \bd x^i \del t) }
$$  -   \hl_{L}^{\hat t} \del \hl^{\hat t}_L +    \l_{L}^i[\delta_{ij} \del  +
F_{ij} (\del  t
+  \ha  A_k\del  x^k)] \l^j_L    + \bar \psi_{R} [ \bd -  ie_0 (A_i   \bd x^i
-\ha F_{ij} \l_L^i\l_L^j) ]\psi_{R}\ \big] \  , $$
where $\l_{L}^{\hat t}  \equiv  \l_{L}^t + A_i \l_{L}^i$. We
have assumed that $T^{IJ}=e_0  \ep^{IJ}, \ I, J=1,2,$
and combined the two Majorana-Weyl spinors $ \psi_{RI}$ into
 a single  Weyl one, $\psi_R=(\psi_{R1} + i \psi_{R2})/\sqrt 2$ (we did
 not write down explicitly  the decoupled  free fermionic terms).
The geometrical  background  corresponding to
 this model is thus the same as  of the
dimensionally reduced bosonic theory \did.
The gauge  invariance   $A'_i=A_i + \del_i \a, \ t'=t - \a, \ \a=\a(x) $
(cf. \gau) is now maintained due to the
  cancellation of the  classical bosonic
 anomaly (in the antisymmetric tensor sector) against the one-loop chiral
fermionic ($\psi_R$) anomaly.

Similarly,  the conformal anomaly which was  previously
absent because of    the contributions
of $\p$ is now  cancelled by the contributions of $\psi_R$
(coming from one  loop order higher).\foot{This  cancellation between different
bosonic and fermionic loop orders  is a direct consequence of the use of
bosonisation and  takes place
also for analogous    heterotic string solutions of  \refs{\gps,\jon}.}
In fact, the leading-order condition of  conformal invariance of this model
 $R_{\m\n}-  \fourth{ \hat H}_{\m\l\r} {\hat H}_\n^{\ \l \r}  - \fourth \a'
{\cal F}(V)^{IJ}_{\m\l}  {\cal F}(V)_{IJ\n}^{\ \ \ \  \l} + ... =0$
is satisfied provided  $e_0^2= 2/\a' $. This is identical to
the bosonic model condition \eee\ with the gauge field term now coming from the
two-loop
fermionic contribution
(the leading terms in the effective actions of the dimensionally reduced
bosonic model and a heterotic model with an abelian vector field are formally
the same if $e_0^2 \sim \a'$).\foot{A simple   indication that such heterotic
model
is,  like the bosonic one, conformally invariant
is that the
contributions of one-loop diagrams with an  internal $t$-propagator
can be cancelled against the contributions of two-loop diagrams  with  the
$t$-propagator replaced by the $\psi_R$-loop
(playing here the  role of the $\p$-propagator of the bosonic model).}
In Section 5.2 we shall give a systematic argument  that \hett\ indeed
represents a conformal heterotic model.

\def \h {\hat }
\def \hi {{\hat i}}
\def \hk{{\hat k}}

\def \hj {{\hat j}}
\def \hu {{\hat u} }
\def \hv {{\hat v} }

%%%%%%%%%%%%%%%%%%%%%%%%%%%%%%%%
\subsec{ Magnetic field from Kaluza-Klein bosonic sector}
%%%%%%%%%%%%%%%%%%%%%%%%%%%%%%%%%%%%%%%%%%%%%%%%%%%%%%%%%%%%%%%
The generalization of \lag\ to the case of the closed superstring theory is
straightforward.
%The superstring  action is given by the
%$(1,1)$ supersymmetric extension of the bosonic $\s$-model
%(with $x^\m= (u,v,x^i)$    in \ replaced by $(1,1)$ superfields $\hat X^\m
%(z, \theta, \bar \theta)$ ).
The  $(1,1)$ supersymmetric \sm action  \ghr\  in the   component
representation is
(here $\hat \omega^m_{\pm n\m} =\omega^m_{\ n\m} \pm \ha H^m_{\ n\m}$ is the
generalised  Lorentz connection, $m, n,...,$ are tangent space indices)
\eqn\onen{I_{(1,1)} = {1\ov \pi \a'}
\int d^2 \s \big[ (G_{\m\n} + B_
{\m\n})(x) \del
 x^\m  \bd  x^\n  + \l_{Rm} [\delta^m_{n}\bd   + {\hat \omega}^m_{ -n\m}(x) \bd
x^\m ]\l^n_R   }  $$
 + \l_{Lm} [\delta^m_{n} \del  + {\hat \omega}^m_{+n\m}(x)\del x^\m]\l^n_L
 - \ha {\hat R}_{+ mnpq} \l^m_L \l^n_L \l^p_R \l^q_R
\big]\ . $$
In the case of the  model \lag\   we get explicitly\foot{The non-vanishing
components of the connection and curvature are:
 $ {\hat \omega}_{-  \hu\hi }=  F_{ij}  dx^j , \
{\hat \omega}_{+  \hi\hj }= - { F}_{ij} du  , \  \ {\hat R}_{+\hj\hk \hu\hi}=
\del_i { F}_{jk} . $ Hats here indicate the
 tangent space indices corresponding to the vierbein $e^\hu=du, \ e^\hv= dv + 2
A_i dx^i, \ e^\hi= dx^i$. The only fermionic vierbein  component   ($\l^{ m}
\equiv e^m_\m \l^\m$)
which is different from the one with a coordinate index  is thus $ \l^\hv =
\l^v + 2 A_i \l^i$ (we  shall consider
$  \l^\hv$ as a new  field replacing  $ \l^v$). }
\eqn\onn{I_{(1,1)} ={  {1\ov \pi \a'} }
\int d^2 \s \big( \del u   \bd  v + 2A_i\del u \bd x^i  +  \del x^i   \bd  x_i
 +  \l_{R}^u \bd \l^\hv_R + \l_{Ri}\bd \l^i_R  } $$
 +  F_{ij} \bd x^j \l_{R}^u\l^i_R
 +   \l^\hv_L \del \l_{L}^u + \l_{Li}\del \l^i_L
 -  F_{ij} \del  u  \l_{L}^i\l^j_L   - \ha \del_i { F}_{jk} \l^j_L \l^k_L
\l^u_R \l^i_R \big)\ . $$
Like the bosonic model \lag, this model is conformal to all loop orders
provided $\del_i F^{ij}=0$ \horts\ (as can be shown, e.g., by repeating the
bosonic argument in terms of $u,v,x^i$ replaced by  superfields).
The  quartic fermionic term in \onn\ is absent in the case of the constant
magnetic field $F_{ij}=\const$.  A simple  test  that this model
 is conformal  follows from the path integral:
if one ignores  the  sources for $v, \ \l^\hv_R$
the integrals over $v$ and $\l^\hv_R$  effectively ``freeze" out $u$ and
$\l_{R}^u$
and thus the interaction terms $ F_{ij} \bd x^j \l_{R}^u\l^i_R  $
and $  F_{ij} \del  u  \l_{L}^i\l^j_L$ in \onen\  do not produce
non-trivial contributions.

Eq. \onn\ can be also interpreted
 as the action of a heterotic \sm  \refs{\sen,\hulw}
corresponding to a ``symmetric"  heterotic  solution
obtained by the standard embedding of a  closed superstring solution into the
heterotic string theory  ($\l_L^\m$ then play the role of the internal fermions
  and $V^{ij}_u={\hat \omega}_{+ u}^{  ij }= - { F}^{ij} $ -- of the internal
gauge field).  This solution
preserves  extended space-time supersymmetry \kallosh.

In addition,
there are two   non-trivial  ``asymmetric"  heterotic  models
corresponding  to $(1,0)$  and $(0,1)$ supersymmetric truncations of \onn\
\horts.
Both  represent {\it  exact } heterotic string solutions  when combined with
a free internal fermionic sector (i.e. there is  no need
to introduce  non-trivial  internal gauge field background).\foot{ The $(1,0)$
truncation also formally preserves space-time supersymmetry and has  extended
world-sheet supersymmetry. This is not surprising given that
 our model with $F_{ij}=\const$  is equivalent (at least in the non-compact
case)  to a non-semisimple WZW model (see Section 2.2). } For example,
 the action for the $(1,0)$ heterotic model reads
\eqn\ohh{I_{(1,0)} ={  {1\ov \pi \a'} }
\int d^2 \s \big[ \del u   \bd  v + 2A_i \del u \bd x^i
  +  \del x^i   \bd  x_i   }
$$  +\  \l_{R}^u \bd \l^\hv_R + \l_{Ri}\bd \l^i_R
 +  F_{ij} \bd x^j \l_{R}^u\l^i_R    +  { \psi}_{LI} \del \psi^{I}_L \big] \ ,
$$
where $\psi^I_L$ are  the  fermions of the internal sector.

%%%%%%%%%%%%%%%%%%%%%%%%%%%%%%%%%%%%%%%%%
\subsec{Magnetic field from the  internal fermionic sector}
%%%%%%%%%%%%%%%%%%%%%%%%%%%%%%%%%%%%%%%%%%%%%%%%%%%%%%%%

\def \B {{\bar B}}
\def \hp {{\hat \p}}
\def \htt { {\hat t}}
\def \h {\hat }
In the above heterotic solutions the magnetic field
came  from the  Kaluza-Klein sector.
Let us now return to the discussion at the beginning of this section and
argue that   the  model \hett\  in which the magnetic field is embedded
into  the internal gauge   sector also represents  an exact
heterotic string solution.
Let us  start  with the
 $(0,1)$ supersymmetric  truncation of \onn\
which, as  explained  above,
is  (for $\del_i F^{ij}=0$)
 an exact conformal model
 \eqn\onk{I_{(0,1)} ={  {1\ov \pi \a'} }
\int d^2 \s \big( \del u \bd v   + 2A_i\del u   \del x^i  +  \del x^i   \bd
x_i   +    \l_{L}^\hv \del \l^u_L + \l_{Li}\del \l^i_L
 -  F_{ij} \del u  \l_{L}^i\l^j_L \big).  }
Changing  the variables to
$\p,t,  \l_L^{\h \p}, \l_L^{\h t}\ $ ($v,u=\p \pm t, \ \l_L^{\hv,u}=
\l_L^{\h\p}\pm \l_L^{{\h t}}, \ \l_L^{{\h t},\h\p}=\l_L^{t,\p} + A_i\l_L^{i},
$\ cf. \la)  and separating  the $\p$-dependent terms we get
\eqn\nnk{I_{(0,1)} ={  {1\ov \pi \a'} }
\int d^2 \s \big[ -\del t \bd t    -  2A_i \del t  \bd x^i  +  \del x^i   \bd
x_i   -   \l_{L}^\htt \del \l^\htt_L   +  \l_{Li}\del \l^i_L  + F_{ij} \del  t
  \l_{L}^i\l^j_L } $$  + \ \l_{L}^\hp \del \l^\hp_L   + \del \p \bd \p + 2( A_i
 \bd x^i - \ha  F_{ij}    \l_{L}^i\l^j_L) \del \p \
 \big]  \ .  $$
The final key step  is to observe that since $\l_L^\hp$ is completely
decoupled,
we can now try to fermionise the compact coordinate $\p$ without breaking
$(0,1)$ world sheet supersymmetry (it will act only on the rest of the fields
excluding  $\l_L^\hp$ and the fermionic counterpart  of $\p$)
 and conformal invariance.
The scheme in which the resulting
model will be conformally invariant  will depend on a  choice of a scheme
used in the fermionisation process.

The action  for a periodic real  boson   $\vp $  ($0<\vp \leq 2\pi$)
\eqn\bose { I_B = {1\ov 2\pi} \int d^2 \s ( \del \vp \bd \vp
+ 2 \B\del \vp + B\B )= {1\ov 2\pi} \int d^2 \s\big[  ( \del \vp + B) ( \bd \vp
+ \B) - F(B) \vp \big] \ ,   }
where $B,\B$ are the components of an   external 2d gauge field ($F(B) \equiv
\del \B - \bd B $),
is equivalent (in the sense of equality
 of the corresponding generating functionals, and, in particular,
the partition functions) to the action for the two Weyl fermions
 (see e.g. \quev\ and refs. there)
\eqn\fermi { I_F = {1\ov 2\pi  } \int d^2 \s
\big[\bar  \psi_{R} (\bd  -i  \B) \psi_R\
+ \bar \psi_{L} \del  \psi_L \big] \  .   }
We are  assuming  that the chiral
 fermionic determinant is defined in the ``left-right decoupled" scheme
in which a specific local counterterm is added
to the ``core"  non-local part $ \sim \int \B ({\del/ \bd}) \B$.\foot{Such
scheme was used also in a similar context   in \gps. In general, the
 bosonisation/fermionisation rule is not unique:
there is
a freedom of adding local counterterms to the non-local part of  the chiral
fermionic determinant. Their choice
depends on symmetry conditions which are  assumed, i.e.  on
definitions of the space-time fields (or couplings  of the theory). For
example, if  one adopts the  ``minimal" (vector-like) scheme  (used, e.g., in
\refs{\quev, \jon})
in which there is no $B\B$ term in \bose\ then  instead of (5.10) below one
ends up with the action
where the target space metric does  not contain the  ``Kaluza-Klein" $A_iA_j$-
term and thus is  not  invariant under the abelian gauge transformations.
The two schemes are related by the field  (metric) redefinition, $G'_{ij}
=G_{ij} + A_iA_j$.
 Such redefinition is also related to  a restoration of  the world-sheet
supersymmetry not apparent
in this ``minimal"  scheme \hult.  Note that in  the present case this
redefinition is not suppressed by an extra power of $\a'$ and  thus the
validity of the use of the second scheme  is not clear.}
Since $\p$ in \nnk\  was     taken  to be periodic with period $2\pi R$,
it can be fermionised {\it provided}  $R^2 = \a'/2 $ (we do not introduce the
Thirring coupling term).
Comparing \nnk\ with \bose\ ($\p= R\vp$)
we learn that $\B=  R\inv ( A_i  \bd x^i -  \ha F_{ij}    \l_{L}^i\l^j_L)$.
Since the supersymmetry should be  present only in the left sector  we should
take
 $B= R\inv  A_i  \del x^i$.
The resulting  conformal
action  is thus (we drop  the
 free $\psi_L, \l_{L}^\hp $-contributions)
\eqn\nnkp{I_{(0,1)}' = {  {1\ov \pi \a'} }
\int d^2 \s \big[ -\del t \bd t    -  2A_i \del t  \bd x^i  +
(\delta_{ij} - A_iA_j) \del x^i   \bd  x^j } $$
 - \   \l_{L}^\htt \del \l^\htt_L +   \l_{Li}\del \l^i_L
  + F_{ij} \del  t    \l_{L}^i\l^j_L  +  \ha F_{ij}  A_k  \del  x^k
\l_{L}^i\l^j_L $$ $$
 +\   \bar \psi_{R} ( \bd -  ie_0 A_i \bd x^i) \psi_{R}
+  \ha i e_0  F_{ij}   \bar \psi_{R} \psi_{R}\l_L^i\l_L^j\ \big]
\   ,  \    \  \   e_0\equiv R\inv = {  \sqrt {2/\a'}}\   .  $$
This is the same as   the $(1,0)$ supersymmetric heterotic action \hett\ which
we have  suggested   above.
 The complete  anomaly-free
heterotic model  is obtained by assuming that $i=1,.., 9$ and adding
to \nnkp\  extra
free 30  Majorana-Weyl fermions in the right sector.

It is clear that the direct supersymmetrisations of our  model
\onn,\ohh,\onk\ can be solved  and quantised  (e.g. using the
 superfield  formulation) in the same way as this was done in the bosonic case
in Sections 3,4. Since the heterotic model \hett,\nnkp\ is closely related to
\onk,
the same should apply to it as well.
The models will be explicitly unitary in the (1,0) superfield version
of light-cone gauge.
Such direct solution will give an explicit definition of the corresponding CFT
(resolving  the  scheme ambiguity in a particular way).

%From the effective action point of view,
%though the leading-order terms in the effective actions of the dimensionally
%%%reduced bosonic  string are formally the same as in the heterotic
%string, higher $\a'$-corrections are different. It would be very interesting
%%%to find the exact $\s $-model backgrounds in the fermionic representation.

%%%%%%%%%%%%%%%%%%%%%%%%%%%%%%%%%%%%%
\newsec{Conclusions }
%%%%%%%%%%%%%%%%%%%%%%%%%%%%%%%%%%%%%%%

  The appearance of tachyons beyond some finite values of the magnetic field,
and the fact that the partition function develops new divergences precisely at
this value, suggest the presence of a phase transition. These are the same
indications that lead  one to think that there is a new, more symmetric string
 phase beyond the Hagedorn temperature \attick.
The present model is exactly solvable and thus may  provide
 a framework to  study explicitly
  possible emergence of such  new symmetries.
 The basic idea is to use the magnetic field as a probe.
 This  can be done, for example,
to reveal  the hidden (spontaneously broken) gauge symmetry   of the
$SU(2)\times U(1)$ GSW model \nielsen.
%For some critical value of the magnetic
%field the ``up" components of the $W$-particles will have zero energy.
%At a
%critical magnetic field where the ``up" components of the W's become
%massless, a W-condensate occurs (but there is still only $U(1)_{em}$
%gauge symmetry). The full symmetry
%is restored  at a higher critical field when the Higgs field becomes
% massless.\foot{We are grateful to P.Olesen for pointing this out to us.}

%At that
%value there is a gauge symmetry corresponding to $SU(2)\times U(1)$
%transformations of these components. It would be very interesting to
%investigate the effective action for  extra zero-energy states that appear at
%special values of the magnetic field.

As we have  have seen  in Section  3,
 the incorporation of a  non-trivial geometry   demanded in the case of
constant magnetic field background by the closed  string  field equations
 has produced
some   substantial  changes  compared to the open string case.
The instability is
   notoriously enhanced: while in the open string theory it takes an infinite
magnetic field
in order to get an infinite number of tachyonic states,
in the closed string theory an arbitrarily small magnetic field already
produces an
 infinite number of tachyonic instabilities.
 It would be interesting to study possible consequences of this  instability in
 quantised
closed string propagation  in  a  magnetic field.
 Above a critical value of the  field,  the tachyonic particles  should
condense, causing a phase transition.
 In string theory this may probably be described by  correlators with multiple
insertion
of   tachyon vertex operators (see also \attick).
%vacuum to produce pairs of tachyonic particles with
%spins oriented in the magnetic field direction.
%The imaginary part of the
%partition function  may be  related to the production rate, and
%may have some semiclassical instanton  interpretation.
%Pair production in the presence of a magnetic field is a sign of a phase
%transition.\foot{It has a different nature than the
%Schwinger pair production in an
%electric field  (a discussion of the
% Schwinger effect in the open string theory was given in \bachas).}
A close
analogue
  is the phase transition that takes place in
type-I superconductors as the magnetic field is increased.
In ref.  \thermal \ a phase diagram
of a closed string gas was obtained
%by demanding finiteness of the thermodynamic partition function
 and  it was argued that the phase transition
 is of  first order  with a large latent heat.
 It would be interesting to investigate
 these issues within the present exactly solvable model.

There are a number of other open problems which deserve  investigation.

$\bullet $ In  Section  5 we have constructed
 several supersymmetric extensions of the
bosonic model. It is important  to compute
 the spectrum and the partition function of these models and to  study
their properties.
%\foot{Since some of these models (but not the  most
%interesting one \nnk)  have residual space-time supersymmetry, the partition
%%function will vanish (unless %supersymmetry is broken by boundary conditions
%%in the compact dimension). }
In the case of the open superstring string theory
in a constant magnetic field  \refs{\abo, \ferrara } the
 spectrum displays the same qualitative features
as in the bosonic case, in particular, the   emergence of the tachyon
instabilities at
 critical values of the field.
 It is plausible that the spectra of the   heterotic string
 models discussed in Section 5
% (in cases when   space-time supersymmetry is broken, e.g.,  by boundary
%conditions in the compact direction)
will also
 be analogous to the bosonic model spectrum, i.e., in particular,
will exhibit similar instabilities.

$\bullet $
It would be important to clarify the implications of
our model for the value of the  $g$-factor in  closed string  theories
and for  possible ``string states -- black holes" connection.
We have seen that the  space-time \did\  describing   response of geometry
to a uniform  magnetic field
is represented by a ``rotating universe" \rotma\
which is not asymptotically flat (cf. \rir).
This  could   imply  a potential ambiguity in the definition of $g$.
The stationary frame used in \did\  leads to conserved string energy
and hence  the corresponding
formula \magmom\ for the $g$-factor derived in our conformally invariant model
seems  applicable to
string states.
One of the consequences of this result is that  a
correspondence between fundamental string states and black holes would become
problematic.
 Indeed, the expression for the thermal average
$ \langle S_R\rangle $ calculated in ref. \rs \ was  $\ha c  \langle S\rangle$,
where $c$
 ($1.27< c <  2$)  depends on the scaling of $S$ with $M$.
The formula \magmom\  has an additional contribution $O(M/Q)$. Therefore,  $g$
will
diverge  unless $M$ scales at most as  $Q$. Since for  black holes the charge
$Q$ scales like the ADM mass $M_{\rm ADM}$,  $\  M$ can scale at most as  $
M_{\rm ADM}$, implying  a conflict in the  counting of states
(the correspondence between level densities requires $M\sim  M_{\rm ADM}^2$
\susskind).\foot {This issue was recently resolved in [57] .}

$\bullet $
 There are at present only very  few  physically interesting  string models  on
non-trivial  backgrounds for  which the
scattering amplitudes are  explicitly calculable.
The present model
should be  one of such examples.
Indeed,  the tachyon vertex operator (see eqs. \soo , \tat \ and \sol )
 % can be written as
%\eqn\tachvett{
%T(u,v,x)=\sum_{l=0}^\infty \int [dp] c_l (e,p)
%e^{ip_uu+ip_v v+ip_1x_1  +ip_ax_a} e^{-\ha \o(x_2-b)^2}H_l[\sqrt{\o }(x_2-b)]
%}
is expressed in terms of exponents of $u,v$ and gaussian exponents and Hermite
polynomials of  the ``transverse"
coordinates $x^i$.
Thus the integrals over $u,v$ can be easily carried out (as was done in the
partition function in Section 4) and then the integrals  over $x^i$ can be
computed  as well  using the  generating functional for $H_l$.

$\bullet $
We have seen that the partition function  of the model \jkl\
is exactly computable as
a function of an  arbitrary magnetic field, i.e.  of free parameters
 $F_{ij}=\const $.
 In contrast to the open string theory where the tree-level (disc)  partition
function  determines the effective (Born-Infeld) action, the  tree-level
(sphere) partition function $Z_0$ in the closed
 string theory is trivial  when  evaluated  at a conformal point  (it vanishes
once divided over  the infinite M\"obius volume). That means that we cannot
use it to get any information about the tree-level  effective action.\foot{
The closed string  effective action $S$
always vanishes when evaluated at a conformal point. This can be
considered as being a consequence of the dilaton equation of motion or  a
consequence of the representation $S = [{\del Z_0 (t)/\del t}]_{t=1}$ where
$t$ is the logarithm of a 2d cutoff, $t=\log \ep$ \tsetl:
 since $Z_0$ computed at a conformal point  is $\ep$-independent,   $S$
vanishes there.}
At the same time, the torus partition function computed as a function of
arbitrary  $F_{ij}=\const $ does give us the value of the
one-loop effective action evaluated on our specific background (with
the square root of the curvature, the   antisymmetric tensor field strength
and the
gauge field strength   all being
proportional to $F_{ij}$), i.e., represents   in certain sense a closed string
analogue of the Heisenberg-Euler-Schwinger  action.
 It may be possible to
use  this expression to obtain  some non-trivial information
about the dependence of the one-loop (heterotic) string
effective action on the gauge field strength.\foot{For example,
expanding $Z$ in powers of $F_{ij}$ one may  compare with
the corresponding low-energy field theory
(e.g. determining a renormalisation of gauge coupling  as in the open string
case  \metsa,  see  also \kkkk).}

$\bullet $ The model \lag\ we have discussed in this paper
is just a simple representative in a more general class of ``chiral null
models" corresponding to  exact string solutions  \refs{\hrt,\horts}.  Other
models in this class describe  plane-fronted waves and
``fundamental string"  type backgrounds  \fund\
 (related, in particular, to extreme electric black holes).
It may be interesting to study if some of our  results can be extended to these
models as well.

$\bullet $ We have seen that the model \jkl\ is closely
related to a  WZW model based on a non-semisimple group.
Though in this paper  we did not  utilise  this connection
  to quantize and solve
the theory,  a  more systematic  current algebra approach  may be useful to
construct    marginal operators and discover hidden  symmetries.

An extension of the present work to a more general class of magnetic
flux-tube backgrounds is discussed in [57].

\phantom{\melvint , \tseta  ,\rutse }

%%%%%%%%%%%%%%%%%%%%%%%%%%%%%%%%%%%%%%%%
\newsec{Acknowledgements}
%%%%%%%%%%%%%%%%%%%%%%%%%%%%%%%%%%%%%%%%%
We would like to thank  G. Gibbons, E. Kiritsis, C. Kounnas, R. Metsaev,
 P. Olesen
 and K. Sfetsos for  useful discussions and comments.
A.A.T. is grateful to Theory Division, CERN for a hospitality during a visit
when this work was at an initial stage and also acknowledges the  support
of PPARC and NATO Grant CRG 940870.

\vfill\eject

\listrefs
\end

\foot{
More precisely,  if one starts with $R\times SU(2)$ WZW  Lagrangian
$L=k(-\del \eta  \bd \eta + \del \psi \bd \psi + \del \theta_L \bd \theta_L
+ \del \theta_R \bd \theta_R  + 2 \cos \psi \del\theta_L \bd \theta_R)$
and takes the limit
$ \psi  = \ha \epsilon  v' + u ,\   \eta   = u ,\
\theta_R= \sqrt \epsilon y_1 , \ \theta_L= \sqrt \epsilon y_2,
k\inv =\epsilon \a', \  \epsilon  \to 0  , $
one obtains \fgf.  This limit does not, however, respect the global structure
of the models: while the Euler angle $\psi$ of $SU(2)$  is periodic,
 $\ha \epsilon  v' + u\to u $ is not periodic in the $E^c_2$ WZW model.
Indeed, if we set
 $t' = \ha (v'-u)$ and  $\phi'= \ha (v'+u)$ (with  $\  v'$ is defined in \coo)
then  $t'= \psi/\epsilon -\eta (\ha + 1/epsilon), \
\phi' = \psi/epsilon + \eta (\ha - 1/epsilon)$  so that both
$t'$ and $\phi'$ are either non-compact or compact at the same time.
To get the model we are interested in (with non-compact $t$ and compact $\phi$)
we may consider the following formal limit:
?????????????????? }

To  check the  correspondence with  the operator quantisation on the cylinder
and to demonstrate  the target space duality invariance
of  $Z$  we need to put   \non\
 into a more suitable form.
Let us first note that \eqn\yyy{
 Y  =  \prod_{(n,n')\not=(0,0)}
[ 1 +   {h\tw\ov n'- \t n}] = \prod_{k\not=0}
[ 1 +  {h\tw\ov k} ] \prod_{ n\not=0, n'}
[ 1 +  {h\tw \ov n'- \t n}] } $$ =
{ \sin (\pi  h \tw)
 \ov  \pi  h \tw}
\prod_{ n=1}^\infty  {(1 - \r\inv
   q^{n })(1 -  \r
   q^{n }) \ov (1 -  q^{n })^{2}} = {1\ov h \tw} {\theta_1(h \tw| \t)
\ov \theta'_1 (0| \t) }  \  ,  $$
where
\eqn\defi{\tw \equiv w'-\t w \ , \ \ \ \  \r\equiv \exp ({2\pi i  h \tw}) \ , \
\ \ q=\exp({ 2\pi i  \t})  \ .
 }